\documentclass[preprint,longnamesfirst]{aastex}
\newcommand{\thisdir}{.}
\newcommand{\kpc}{\mathop{\rm kpc\,}\nolimits}
\newcommand{\Mpc}{\mathop{\rm Mpc\,}\nolimits}
\newcommand{\kps}{\mathop{\rm km/s\,}\nolimits}
\newcommand{\Mach}{\mathop{\mathcal{M\,}\,}\nolimits}
\newcommand{\dyne}{\mathop{\rm dyne\,}\nolimits}
\newcommand{\K}{\mathop{\rm K\,}\nolimits}
\newcommand{\Msun}{\mathop{\rm M_{\odot}\,}\nolimits}

\newcommand{\fig}{Fig.~\ref}
\newcommand{\figs}{Figures~\ref}
\newcommand{\Fig}{Figure~\ref}
\newcommand{\tab}{Table~\ref}
\newcommand{\tabs}{Tables~\ref}
\newcommand{\sect}{Sec.~\ref}

\newcommand{\eq}{Eq.~\ref}
\newcommand{\Hydra}{{\sc hydra}}

\newcommand{\expd}[1]{\times 10^{#1}}
\newcommand{\mean}[1]{\langle #1 \rangle}
\newcommand{\vect}[1]{\mathop{\bf #1\,}\nolimits}
\newcommand{\grad}{\nabla}
\newcommand{\divdotv}{\grad\cdot\vect{v}}

\newcommand{\fraction}[2]{\mbox{\scriptsize$^{{#1}\!}/_{\!{#2}}$}}

\ifx\undefined\onequarter
 \newcommand{\onequarter}{\fraction{1}{4}}
\else
 \renewcommand{\onequarter}{\fraction{1}{4}}
\fi
\ifx\undefined\onethird
 \newcommand{\onethird}{\fraction{1}{3}}
\else
 \renewcommand{\onethird}{\fraction{1}{3}}
\fi
\ifx\undefined\onehalf
 \newcommand{\onehalf}{\fraction{1}{2}}
\else
 \renewcommand{\onehalf}{\fraction{1}{2}}
\fi
\ifx\undefined\twothirds
 \newcommand{\twothirds}{\fraction{2}{3}}
\else
 \renewcommand{\twothirds}{\fraction{2}{3}}
\fi
\ifx\undefined\threequarters
 \newcommand{\threequarters}{\fraction{3}{4}}
\else
 \renewcommand{\threequarters}{\fraction{3}{4}}
\fi
\newcommand{\threefourths}{\threequarters}

\newcommand{\figscale}{1}
\newcommand{\figscaletwo}{.4}

\begin{document}

\title{Hydrodynamical Drag in Cosmological Simulations}
\author{Eric R. Tittley and H. M. P. Couchman\footnote{Present address: McMaster University, Department of Physics and Astronomy, Hamilton, Ontario, L8S 4M1, Canada}}
\email{etittley@astro.uwo.ca}
\email{couchman@physics.mcmaster.ca}
\affil{University of Western Ontario}
\affil{
Department of Physics and Astronomy, 
London, Ontario, N6A 3K7, 
Canada }
\author{F. R. Pearce}
\affil{
Department of Physics, University of Durham, \\
Durham, DH1 3LE, United Kingdom
}
\email{F.R.Pearce@durham.ac.uk}

\begin{abstract} 

We present a study of hydrodynamic drag forces in smoothed particle simulations. In particular, the deceleration of a resolution-limited cold clump of gas moving through a hot medium is examined.  It is found that the drag for subsonic velocities exceeds that predicted by simple physical approximations.  This is shown to be a result of the hydrodynamical method which encourages the accretion of particles from the hot medium onto a shell around the cold clump, effectively increasing the radius of the clump.  For sonic and supersonic velocities, the drag is shown to be dependent on the effective cross section of the clump.  The consequences for cosmological simulations are discussed.
\end{abstract}

\section{INTRODUCTION}
\label{sec.Introduction}

Numerical simulation has proved to be a powerful tool for understanding the dynamics and formation history of cosmological structure \citep{DEFW92,Katz92,Cen94,EMN}.  When N-body techniques are combined with a hydrodynamical method to follow the baryonic component, further advancements can be made as now no additional assumptions are required to relate the baryonic material to the underlying collisionless matter.  Such a combination has proved to be both a popular and powerful technique \citep{HK89,Evrard90,Katz91,Cen92,Navarro93,SM93,ROKC93,Bryan94,Gnedin95,Steinmetz96}.  A common method of implementing hydrodynamics in N-body simulations is through the use of smoothed particle hydrodynamics (SPH) \citep{GM77,Lucy77}.  The great strength of SPH is that it is a Lagrangian technique; individual particles act as tracers of the flow and better resolution
can therefore be achieved in regions of high tracer density. This is a very desirable property for cosmological simulations where extremely large density contrasts are present and excellent spatial resolution needs to be achieved within relatively large volumes. 

In the standard model of cosmological structure formation, larger structures are formed via the amalgamation of smaller structures formed at earlier times in a hierarchical fashion.  In any simulation modelling the formation of such structures in this hierarchical manner, there will be an abundance of objects formed at the resolution limit of the simulation. A particularly difficult problem is following these small, resolution-limited halos as they are subsumed into larger structures as the hierarchy progresses.  It has been found that much of the substructure in collisionless simulations is erased unless very high resolution is employed \citep{Ghigna98,Moore99,KGK99}.  This is the classic overmerging problem \citep{White76,vK95,SDE95,Moore96}.

With the inclusion of radiative cooling into the hydrodynamical model, the baryonic material can dissipate energy and become more tightly bound than the dark matter, allowing individual objects to survive the hierarchical process that builds up large structures \citep{SDE95,FEWS}.  \citet{KGK99} argues that these objects are not likely representative of bound gaseous structures in nature, since they are lacking the dark matter halos observed to dominate similar structures \citep{PSS96}.  As well, the dynamics of this set of objects may now be affected by gas forces; as \citet{FEWS} demonstrates, the viscous drag due to the passage of a galaxy-size object through the hot intracluster medium can be sufficient to cause much of the material to spiral into the centre.

An understanding of the numerical and physical processes involved as resolution-limited gaseous structures move around within the environment of larger halos is essential to understanding the results of numerical simulations.  This includes the effect on the global properties of the larger halos.  Velocity bias, where galaxies have a smaller velocity dispersion than that predicted by virial arguments due to the exchange of orbital energy to the collisionless material is sometimes observed in simulations \citep{ESD94,FEWS} and sometimes not \citep{KHW92} (However, \citet{SDE95} claims these results are strongly affected by the choice of galaxy tracer).  Spatial bias, in which the galactic population is more or less centrally concentrated than the dark matter, again due to exchanging orbital energy with the collisionless material, is another effect sometimes observed in simulations of the formation of galaxy clusters \citep{CO93a}. This can lead to incorrect estimates of the underlying density profile.  Observationally, velocity and spatial bias have serious effects on mass estimates for clusters \citep{FG83,JF84,Girardi98} and hence on such fundamental quantities as the baryon density and the amount of matter in the Universe as a whole \citep{WF95,WJF97}.  Mass segregation, in which larger galaxies are observed to lie, on average, closer to the centre of clusters than smaller galaxies, is another effect which occurs in many models of cluster formation \citep{FEWS}.  An understanding of the processes that affect both the rate by which the smaller structure is merged into larger halos and the final location of the deposited material are of fundamental importance.

Any process which modifies the merger rate of small groups of galaxies in simulations also impacts the connection with observations and models of galaxy populations.  It may have an important bearing on the physical processes that lead to the formation of giant cD galaxies. In addition, there has been much interest in the morphology-density relation in which elliptical galaxies are observed to be much more common in rich environments than spirals \citep{Dressler80}, a result supported by simulations \citep{EYKK99,CO93b}.  The merger of spirals to form ellipticals is currently a very popular model \citep{WR78,Cole91,LS91,WF91} and the rate of this process will be determined by the detailed dynamics of galaxies within clusters. The merger rate of galaxies within halos is also an important parameter for the semi-analytic models of galaxy formation \citep{KWG93,Cole94,Baugh98}. These rates are often taken from low resolution SPH simulations similar to those we discuss here.

The effect of drag on clumps of matter is significant in simulations of other astrophysical scenarios.  SPH has been used to simulate the feeding of active galactic nuclei \citep{AL96,SM94}, starbursting behaviour within massive galaxies \citep{HS94}, formation of interstellar clouds through the merger of clumps \citep{Kamaya97,BFWW98}, and, more recently, even planet formation \citep{NBAA98}, all situations in which the merger rate of small objects is of vital importance.

In \citet{Thacker98}, tests of a variety of implementations of SPH are described.  The implementations varied in the method of force symmetrisation, the form of the artificial viscosity, and the presence of a shear-correction term in the artificial viscosity.  They were compared by probing the behaviour of the SPH implementations in a series of scenarios relevant to cosmological simulations.  Included in this series is the drag experienced by a cold clump travelling through a hot medium.  It was concluded that the drag on the cold clumps is excessive when SPH is used, independent of the implementation method.  The reason for this behaviour was not explored, nor was the behaviour extensively characterised. This paper is an extension of the drag tests work described in \citeauthor{Thacker98}.  We present here the 

In this paper we investigate the numerical processes which control the strength of the viscous drag of a hot medium on a resolution-limited cold clump of gas within a smoothed particle hydrodynamical (SPH) method.  This is done through the use of numerical simulations of the simple scenario in which a cold gaseous halo is passed through a tenuous hot, uniform medium.  Three velocity regimes are explored: subsonic, sonic, and supersonic. Analysis is assisted by a comparison of the actual deceleration observed for our clumps with that predicted from simple theoretical arguments.

Our model clumps are intentionally simple. We study what are essentially resolution dominated spheres of material moving within a uniform density hot medium.  Such objects are typical of the small galaxies or groups of galaxies formed in cosmological simulations of structure formation where most of the objects have masses and physical extents near to the resolution limit of the model \citep{Pearce99}.  It is important to note that regardless of the resolution, these resolution-limited structures will form and play a role in the development of the larger structures.

The paper is laid out as follows. In \sect{sec.SPH} we include a brief overview of the relevant SPH equations highlighting those factors which affect the drag.  The model clump-halo systems are described in \sect{sec.tests.drag.systems}.  The predicted deceleration for a galaxy which sweeps up all the material it encounters is derived in \sect{sec.Expectation}.  A comparison between this and the actual deceleration found in our models for a variety of Mach numbers is presented in \sect{sec.Results} and discussed in \sect{sec.Discussion}.

\section{SMOOTHED PARTICLE HYDRODYNAMICS}
\label{sec.SPH}
\renewcommand{\thisdir}{SPH}
\subsection{Force symmetry}
Under the SPH formalism any scalar quantity at any position can be calculated
by averaging the properties of the surrounding particles. This average  is generated by weighting the contribution of each particle as a function of the distance from the designated point using a weighting function (or kernel) that is smoothly varying.

The average of any scalar quantity, $A(\vect{r})$, is given as;
\begin{equation}
\mean{A(\vect{r})} = \int d\vect{r}' A(\vect{r}') W(\vect{r}-\vect{r}',h),
\label{eq.tests.intro.SPHmean_cont}
\end{equation}
where $W(\vect{r},h)$ is the kernel function and $h$ is the smoothing
length.  The kernel function is generally, but not necessarily,
spherically symmetric, has compact support and is normalised such
that its spatial integral is unity. 

The gradient of any scalar field can be calculated by using the
gradient of the kernel function.  Noting that the integral becomes a summation over $N$ neighbouring particles for a finite distribution of particles, the gradient estimate is,
\begin{equation}
\grad\mean{A}_i = \sum_j^N  \frac{A(\vect{r}_j)}{n_j} \grad W(\vect{r}_i-\vect{r}_j,h),
\label{eq.tests.intro.SPH_grad}
\end{equation}
where $n_j$ is the local particle number density.

In the above description, the smoothing length, $h$, has been taken to be a constant.  However, to make the best use of the Lagrangian nature of the technique, this would be inappropriate since it would establish a minimum resolution scale far larger than the resolution permitted by the increased number density of particles in the dense regions.  To allow for this, the smoothing length should decrease as the local number density increases to maintain an appropriate spatial resolution.  For $h$ to scale appropriately with the resolution it is normally taken to be a distance which encompasses some constant number of neighbouring particles, $N_{SPH}$.

Allowing $h$ to vary leads to the question of which particle's smoothing length should be used in the SPH summations.  There are two interpretations of SPH; the `gather' and the `scatter'.  In the `gather' interpretation, a local estimate of a property about a particle, $i$, is made by sampling out to some fixed distance and weighting each particle found by its separation from the centre, hence the smoothing length in the summation is $h_i$. In the `scatter' interpretation a local estimate is made using the sum of the contributions to that position from all the particles for which their smoothing length extends to encompass the position.  Here, $h_j$ is used.  In practice, neither interpretation is accepted but rather the appropriate smoothing length is some function of the individual $h$'s, $h_{ij}=f(h_i,h_j)$.  The form of $h_{ij}$ can be the arithmetic, harmonic or geometric mean of the $h$'s, or some other form.  If it is symmetrical, then $f(h_i,h_j)=f(h_j,h_i)$. It need not be symmetric in the general case (for example, to calculate densities), but if it is to be used in the calculation of the force, symmetrisation will preserve force symmetry and conserve momentum.  A further, equally valid, option is to average the kernels themselves; $W(\vect{r}_i-\vect{r}_j,h)$ is replaced with $\left( W(\vect{r}_i-\vect{r}_j,h_i)+W(\vect{r}_i-\vect{r}_j,h_j) \right)/2$.  This also preserves the force symmetry and conserves momentum.

To demonstrate the utility of these symmetrisation methods, let us examine the basic hydrodynamical force equation. The hydrodynamical component of the force, $F$, on each particle comes from the pressure gradient, $\grad P$, via
\begin{equation}
F=-m\frac{\grad P}{\rho},
\label{eq.tests.intro.PressureGrad}
\end{equation}
where $m$ and $\rho$ are the gas particle mass and density, respectively.  To preserve force symmetry it is necessary to find an SPH approximation of the hydrodynamical force contribution such that $F_{ij}$ = $-F_{ji}$ where $F_{ij}$ is the force felt by particle $i$ due to particle $j$.  This is frequently achieved by using the identity;
\begin{equation}
\frac{\grad P}{\rho} = \grad\frac{P}{\rho} + \frac{P}{\rho^2}\grad{\rho}.
\label{eq.tests.intro.Identity}
\end{equation}
Combining this with \eq{eq.tests.intro.PressureGrad} we get,
\begin{equation}
  F_{i} = -m_i m_j\left(\sum_j^N\frac{P_j}{\rho_j^2}\grad W(\vect{r}_i-\vect{r}_j,h_j) +\sum_j^N \frac{P_i}{\rho_i^2}\grad W(\vect{r}_i-\vect{r}_j,h_j)\right).
\label{eq.tests.intro.ForceCalc1}
\end{equation}
This is not yet symmetric.  When the force on particle $j$ is computed, the contribution from particle $i$, $F_{ji}$, will not be the negative of the contribution $F_{ij}$ to $F_i$ since $\grad W(\vect{r}_i-\vect{r}_j,h_j)
\neq -\grad W(\vect{r}_j-\vect{r}_i,h_i)$ if $h_i \neq h_j$.  However, if $h_i$ and $h_j$ are replaced with a symmetric combination, $h_{ij}$, it is clear that the gradient will also be symmetric.  For a spatially varying smoothing length, \citet{Monaghan92} suggests taking the gradient of some average of the kernels calculated symmetrically;
\begin{equation}
F_{i} = -m_i m_j\sum_j^N \left(\frac{P_j}{\rho_j^2}+\frac{P_i}{\rho_i^2}\right) \grad \bar{W}_{ij}.
\label{eq.tests.intro.ForceCalcSM93}
\end{equation}
The `average' kernel, $\bar{W}_{ij}$, can use the kernel averaging
scheme seen above or some average of the smoothing lengths for
$h_{ij}$.  That is, either
\begin{equation}
\grad \bar{W}_{ij} = \left[\grad W(\vect{r}_i-\vect{r}_j,h_j) + \grad W(\vect{r}_i-\vect{r}_j,h_i)\right]/2
\end{equation}
or $\grad \bar{W}_{ij} = \grad W(\vect{r}_i-\vect{r}_j,h_{ij})$ where $h_{ij}$ is an average of the smoothing lengths seen before. The former approach is used in \citet{SM93}.

\begin{figure}
%\epsscale{\figscale}
\plotone{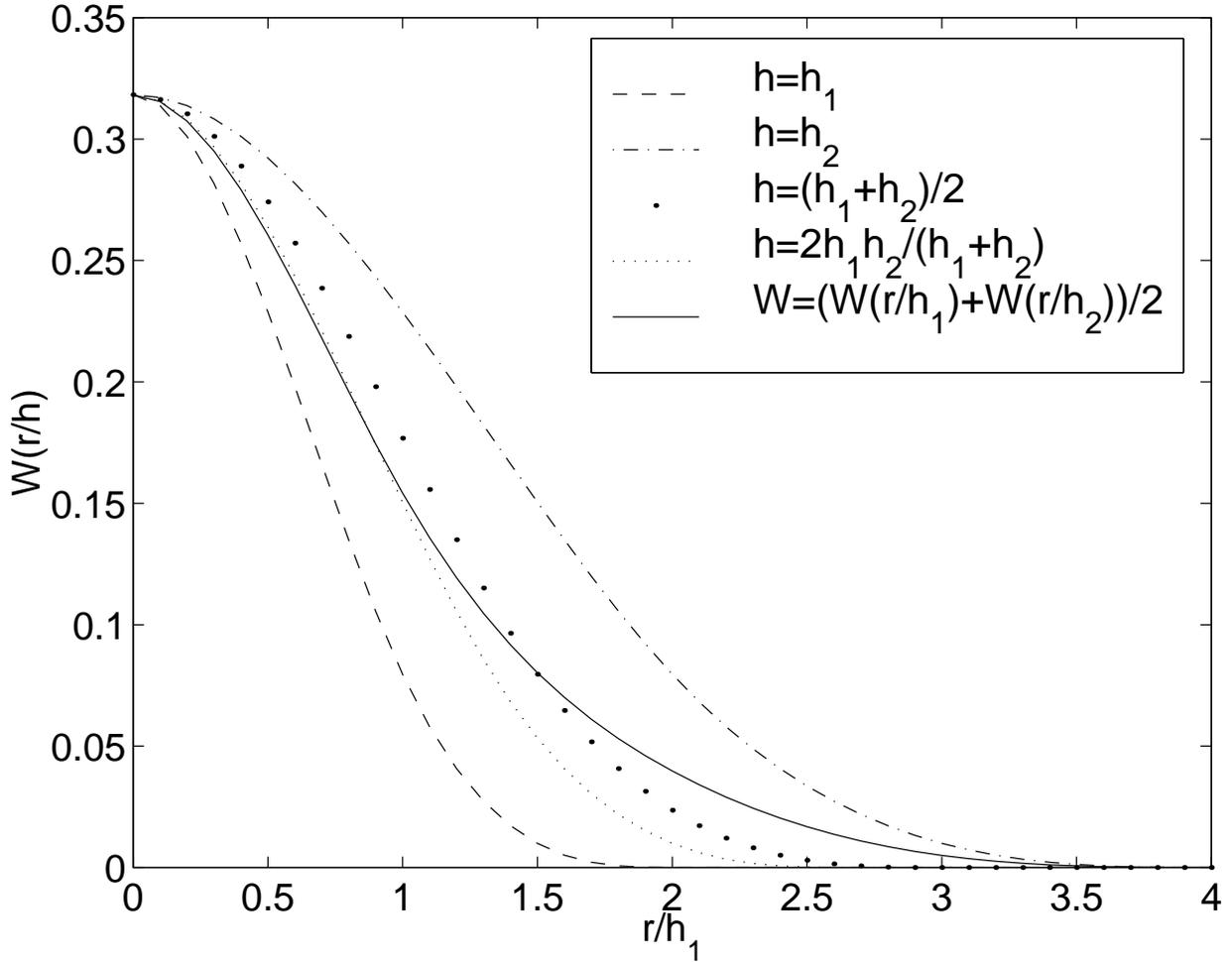}
\caption{The kernel function for differing effective smoothing lengths.
Plotted are the kernel functions for the symmetrisations of $h$ via the arithmetic mean (large dots) and harmonic mean (small dots) when $h_2=2h_1$.  The kernel functions using just $h_1$ and $h_2$ are also given (dashed and dot-dashed, respectively).  The effective kernel function for the average of these two kernels is also plotted (solid line).
}
\label{fig.SPH.KernelShapes}
\end{figure}

The effective kernel shapes produced by two smoothing radii that differ by a factor of two are illustrated in \fig{fig.SPH.KernelShapes}. The effective kernel shapes differ most appreciably in the wing of the function, particularly between the $h$-averaging and the kernel averaging schemes.  For the massive cold clump, the increase in the wing for the kernel averaging scheme increases the effective cross-section of the clump and, consequently, increases the drag (see \sect{sec.Results.Drag}).

\subsection{Artificial viscosity}
Early on, it was realised that for SPH a viscosity force term would be necessary to damp the flow of gas (see, for example, \citet{Lucy77}).  This is entirely to be expected; the gas is represented in the simulation by particles of macroscopic size, while viscosity is due to turbulence on all scales, including microscopic.  The cascading of energy from large-scale motions to small-scale turbulence and eventually, via atomic interactions, into heat must be modelled by a viscosity term in the hydrodynamic equations.  The adopted forms of the viscosity terms are essentially {\it ad hoc}.  They are all designed to cover bulk viscosity and shock front dissipation \citep{Monaghan92}.  In some cases, it is important to modify these forms to permit shear \citep{Balsara}.  All, however, convert convergent kinetic energy into thermal energy.

It will be shown in \sect{sec.Results.Subsonic} that the triggering of the artificial viscosity does affect the drag felt by the clump in the subsonic case.  However, the form of the artificial viscosity is not significant to the general behaviour of the drag, as is shown in \citet{Thacker98}.  Interestingly, they report that artificial viscosity actually decreases the amount of drag experienced by cold clumps, since it reduces the penetration of hot particles into the cold clump.  Without the artificial viscosity, the convergent flows experienced ahead of the clump are not damped, allowing particles to pass through the bow wave and approach closer to the cold clump.  This increases the instantaneous pressure force between the hot particle and the cold clump leading to a collision more elastic in nature.  Elastic collisions take away up to twice as much momentum as inelastic.  With the artificial viscosity term, the hot particles are diverted into the flow around the clump, heading off any close approaches.

\section{NUMERICAL METHOD}
The implementation of SPH used here was kindly made available by Dr. R. Thacker and is described in \citet{Thacker98}.  It is based on the N-body hydrodynamic code, \Hydra\ \citep{CTP}.  Further details about the code are available there.  A more complete description of the methods for generating the symmetrical hydrodynamical forces are contained in \citet{Thacker98} as well as the implementation of the form viscosity. We use a Monaghan-type artificial viscosity.  Symmetrisation of the force equation is provided by arithmetic averaging of the smoothing lengths unless otherwise noted.

Since it was demonstrated in \citet{Thacker98} that the general behaviour of the drag is not affected by the presence of a Balsara term \citep{Balsara}, which reduces the viscosity in the presence of shear, one was not included in the implementation of SPH used.

For the calculation of the gravitational forces, \Hydra\ uses an adaptive particle-particle, particle-mesh (AP$^3$M) method.  The minimum resolution is set by the gravitational softening length which is $75\kpc$.

All forces are periodic across the boundaries of the simulation cube.

\section{MODEL SYSTEMS}
\label{sec.tests.drag.systems}
\renewcommand{\thisdir}{Simulations}
The tests in this paper are all based on a model containing a cold clump moving through a hot medium.  This clump may be allowed to slow or be forced to maintain a constant velocity, depending on the needs of the particular test. Our models attempt to mimic the conditions encountered as a resolution-limited cold clump passes through the halo of a larger structure.
The models consist of a knot of 100 cold particles moving through a uniform distribution of 13000 hotter particles.  The volume is $2\Mpc$ on a side and has periodic boundary conditions.

The clumps of cold gas are essentially structureless spheres, lying at the resolution limit of the code since they are to represent the first (and most common) structures to form in a cosmological simulation of hierarchical clustering which is the dominant source of material during the creation of larger structures \citep{Kay99}.  The gas in the clumps is cold ($10^4 \K$) since the dense gas cools rapidly to this temperature through a radiative process.

Within a typical cosmological simulation the halos of galaxy clusters are hot ($\la 10^8 \K$) and more diffuse than the cold clumps because the hot halo gas is sitting close to virial equilibrium within the cluster potential well. This hot halo is observed through the X-ray emission (see \citet{HMS99}, for examples).  Consequently, the hot media in our models are at temperatures of $10^7$ or $10^8 \K$ and are $1/100$ as dense as the cold clumps.

To cover a variety of infall speeds, we examine the deceleration of a knot of cold gas in three velocity regimes: Mach 2, Mach 1, and Mach \onethird. The Mach 2 and Mach 1 tests differ in terms of the speed of the cold knot; the temperature of the hot gas is kept fixed.  The Mach \onethird\ test uses the same clump velocity as the Mach 1 test, but is performed in hotter gas. \tab{Tab.Drag.Init} gives the details of the cold clump and hot gas phases. 

The hot gas is in a glass configuration which was prepared from an initially random placement of particles and subsequently allowed to relax to a stable state. The cold clump was created by randomly placing stationary particles within a sphere of radius equal to the gravitational softening length then allowed to relax in isolation.  The two systems were then combined.

The Jeans length, $R_J$, for the hot gas phases is sufficiently large to ensure stability even in the presence of the perturbation from the cold clump. Consequently, dynamical friction should not be important.  This conclusion was confirmed by passing a collisionless clump through the hot medium -- it experienced negligible deceleration.  \citet{Ostriker99} discusses dynamical friction in systems with hydrodynamics. 

The box length, $2\Mpc$, was chosen so that the cold clump was well separated from its images (arising from the periodic boundary conditions employed) and would move across the box only once without encountering its own wake.

\section{EXPECTED DECELERATION}
\label{sec.Expectation}
The expected maximal rate of deceleration can be approximated by considering a disc sweeping through a medium, collecting all the material it encounters. This would represent the maximum expected rate of deceleration if dynamical friction and elastic collisions are ignored. From the equations of conservation of mass and momentum, the solution for the velocity, $V$, of such a system is found to be
\begin{equation}
V(t)=V_o\left(\frac{t}{t_{\onehalf}}+1\right)^{-\onehalf},
\label{eq.Expected}
\end{equation}
where $t_{\onehalf}$ is a characteristic time scale corresponding to the time at which a mass of the hot gas equal to one half that of the disc has been swept up.  It is given by
\begin{equation}
t_{\onehalf}=\frac{M}{V_o 2\pi R^2 \rho_g}.
\label{eq.tonehalf}
\end{equation}
Here, $M$ is the mass of the disc at the start, $R$ is the radius of the disc, and $\rho_g$ is the density of the gas through which the disc is travelling. The disc starts with velocity $V_o$ at time $t=0$. The expected acceleration is just the derivative of this expression, and at $t=0$,
\begin{equation}
\frac{dV}{dt}=-V_o^2\frac{\pi R^2\rho_g}{M}.
\label{eq.expected.drag}
\end{equation}

\citet{RS71} derives a similar expression for the contribution from bow shock drag to the deceleration of a galaxy passing through an intracluster medium.  Its form differs in that it contains a factor for the coefficient of drag, $C_D$, which is approximately unity.

For our models the values of $M$, $V_o$ and $\rho_g$ are given.  For any decelerating system, the drag may be associated with \eq{eq.expected.drag}, allowing the calculation of an effective radius, $R_{eff}$.  An estimate of $R_{eff}$ may be made as follows.  For the clump particles, the smoothing length, $h_{clump}$, is set to the minimum allowable value, as determined by the gravitational softening length.  This leads us to choose $R_{eff}=2h_{clump}$. However, because of the symmetrisation inherent in the SPH method, the smoothing length of the hot particles is also important and it is some combination of this
length with the smoothing length of the cold particles that is required.  We will assume that the appropriate radius is that radius at which the clump particles contribute an equivalent amount to the SPH density calculation for a hot gas particle as the hot particle itself. That is,
\begin{equation}
N_{clump} W(R_{eff},h_{ij}) = W(0,h_{hot}).
\label{eq.ReffEst}
\end{equation}
Since $N_{clump} >> W(0,h_{hot})$, this implies $W(R_{eff},h_{ij}) \simeq 0$ which gives $R_{eff} \simeq 2h_{ij}$.

\renewcommand{\thisdir}{Results}
\section{RESULTS}
\label{sec.Results}
Using the model clump-hot-gas systems, a series of tests were performed to characterise the behaviour of the drag in three velocity regimes: subsonic, sonic, and supersonic.  A description of the individual tests as well as their results are presented in the following subsections. The results of these tests are broken into those that pertain to the clumps in all of the velocity regimes and those that are peculiar to the subsonic regime.

\subsection{Viscous drag in three velocity regimes}
\label{sec.Results.Drag}
The behaviour of the drag that is common to the velocities spanning subsonic to supersonic is described here.  First, we will demonstrate that the viscous drag for subsonic clumps is much greater than that expected from theoretical arguments.  For all velocity regimes, it will then be shown that the effective radius of the clump, as determined by the drag it encounters, scales with the averaged smoothing length, $h_{ij}$.  It will be verified that as a consequence, the drag varies with the method of force symmetrisation.  Since $h_{ij}$ is also a function of number density contrast, the increase in drag in regions of higher density will be demonstrated to be partially offset by the decrease in the averaged smoothing length.  The distribution of the energy deposition will also be examined.

\subsubsection{The measured drag {\em versus} the fiducial drag}
\label{sec.Results.Drag.Excess}
The fiducial drag, given by the analytic expression \eq{eq.expected.drag} amounts to a {\em maximum} amount of expected deceleration, presuming inelastic collisions and a non-rigid medium. A measured drag which exceeds the corresponding fiducial value for the clump would necessarily by excessive.  Measurements of the drag on the clump as it travels at a variety of velocities compared with the fiducial amount delineate those velocities for which the drag is excessive.  To this end, a series of tests was conducted which allowed the drag to be calculated for such clumps.

\begin{figure}
\epsscale{\figscale}
\plotone{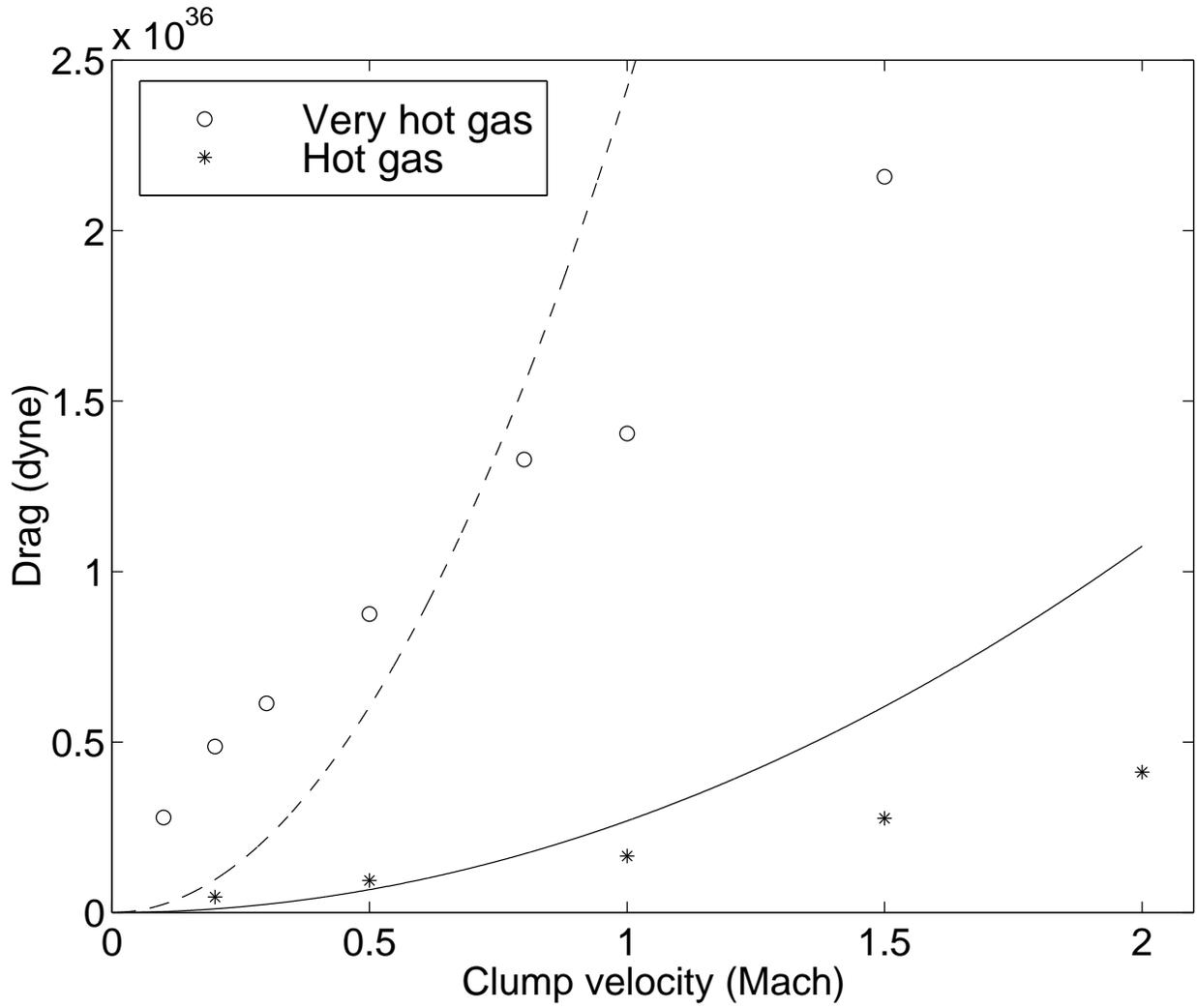}
\caption{The mean drag felt on a clump moving with constant velocity.
The data are given for both the very hot gas ($V_s = 1500 \kps$) and
the hot gas ($V_s = 500 \kps$). The lines indicate the theoretical
maximum deceleration calculated in \sect{sec.Expectation}.}
\label{fig.Results.MeanClumpDrag}
\end{figure}

The tests involved ploughing a clump through a uniform medium at various fixed velocities.  The mean drag on the clump was measured from the force required to keep the clump moving at a constant speed.  The drag, as it varies with Mach number, is illustrated in \fig{fig.Results.MeanClumpDrag}. The results indicate a linear trend of the drag with velocity for speeds greater than about Mach~\onehalf.

If the fiducial force of drag given by \eq{eq.expected.drag} is assumed and the radius of the clump, $R$, is taken to be twice the combined smoothing length, $h_{ij}$, as detailed in \sect{sec.Expectation}, then any points that lie above their respective curves in \fig{fig.Results.MeanClumpDrag} indicate Mach numbers for which the drag on the clump is excessive.  This is clearly the case for velocities below Mach~1.  The Mach number of the transition to the domain of excessive drag is similar regardless of the sound speed for these tests, despite the fact that the speed of sound varies by a factor of three.

The results may be illustrated by another method.  The measured drag force equated with the left hand side of \eq{eq.expected.drag} gives an effective radius, $R_{eff}$.  This radius, and consequently the effective cross section of the clump, is much larger than the estimated value of $2h_{ij}$ at slow speeds, which is shown in  \fig{fig.Results.ReffVsMach}.  Indeed, it is larger than $2h_{hot}$, the maximum sphere of direct influence possible through SPH force calculations.

At supersonic speeds there is an apparent convergence of the effective radius to a value less than the estimated $2h_{ij}$.  Indeed, it seems to be converging to $h_{ij}$, which is not surprising since the density of the bow shock will increase as the clump velocity rises while \eq{eq.ReffEst} assumes the hot particles are relatively isolated compared with the density of the clump.  However, the relation between the effective radius and the Mach number, $\Mach$,
\begin{equation}
\frac{R_{eff}}{h_{ij}} = 1.66\pm 0.08 \Mach^{-0.59\pm 0.04},
\label{eq.ReffFit}
\end{equation}
illustrated in \fig{fig.Results.ReffVsMach} also fits well and it converges to zero for large $\Mach$.  Thus, there is not enough information to argue for a convergence to any particular value.

The effective radius of the clump may be made consistent with the estimated value of $2h_{ij}$ if we include a coefficient of drag in \eq{eq.expected.drag} with a value $C_D = 0.5$.

It is notable that the effective radius is not a function of the speed itself, but of the speed scaled by the local speed of sound.  This emphasizes the velocity scale invariance of the results presented in this paper.

The excessive drag at subsonic speeds manifests itself in \fig{fig.Results.MeanClumpDrag} by the excessive drag over the fiducial force of drag and in \fig{fig.Results.ReffVsMach} as the rapidly increasing effective radius towards lower speeds.  We will return to this point in \sect{sec.Results.Subsonic} where the cause of this phenomenon will be explored.

\subsubsection{The scaling of the drag with the averaged smoothing length}
\label{sec.Results.Drag.h_ij_Scaling}
If the effective radius of the clump is truly set by the radius of the combined smoothing length, $h_{ij}$, as is argued in \sect{sec.Expectation}, then the drag on the clump should scale with this length.  To test this the clump was passed through three media of the same physical density, but differing number density, $n_{hot}$, of hot gas particles. Since the smoothing length, $h_{hot}$, scales with $n_{hot}^{\onethird}$, this varies the mean smoothing length of the hot gas and consequently the combined smoothing length, $h_{ij}$. The test was done for number density increments of two and five times. In these tests, which were performed for the velocity regimes of Mach~\onethird, 1, and 2, the clumps were permitted to slow instead of being held at constant velocity as was done for the tests in the previous section.

For each of the tests, an effective radius, $R_{eff}$, was determined by fitting the velocity of the clumps to the form of \eq{eq.Expected}.  The results, given in \tab{tab.Results.VaryingNden}, support the hypothesis that the effective radius does indeed scale with the averaged smoothing length, $h_{ij}$.  In each velocity regime, the effective radius is approximately a constant multiple of the averaged smoothing length (fifth column).  The correlation is not strong enough to be distinguished from scaling with just the smoothing length of the hot gas, $h_{hot}$ (fourth column).  It is clear, however, that the effective radius does not scale with the smoothing length of the clump particles (third column).

Though the effective radius is a similar factor of the averaged smoothing length within each regime, the factor is different between each regime; $R_{eff} \simeq 3$ for the Mach~\onethird\ tests, $2$ and $1.5$ for the Mach~1 and Mach~2 regimes, respectively.  This trend is the same as outlined in \sect{sec.Results.Drag.Excess} and, in particular, \fig{fig.Results.ReffVsMach}. These scaling coefficients may be compared with the estimated factor of $2$ made in \sect{sec.Expectation}. Only the deceleration of the Mach~1 clump matches, while the effective radius of the supersonic clump is slightly less than the estimate.  The result for the Mach~\onethird\ clumps further emphasizes the excessive amount of drag found in the subsonic regime.

\subsubsection{The scaling of the effective cross section with local density}
\label{sec.Results.Drag.Rho_scaling}
The results of the drag tests involving the deceleration of a clump through media of varying number density (\sect{sec.Results.Drag.h_ij_Scaling}) indicate that the drag felt on a clump travelling in the outer halo of a gravitationally bound system will be increased due to the low number density of particles in this part of the halo.  As the clump travels into denser parts of the halo, the apparent cross section will be reduced due solely to the increased number density of the surrounding particles.  This should offset to a degree the increase in drag that the clump would be expected to feel due to the increase in the physical density.  The force of drag in \eq{eq.expected.drag} is related to the effective radius and the local gas density via $F_{drag} \propto - R_{eff}^2 \rho_{hot}$.  Since we have the relations, $R_{eff} \propto h_{ij} \propto n_{hot}^{-\onethird}$, then $R_{eff} \propto \rho_{hot}^{-\onethird}$ may be derived.  This implies that the drag force for the clump scales with the local gas density as $F_{drag} \propto - \rho_{hot}^{\onethird}$.

This result assumes $h_{ij} \propto h_{hot}$, an approximation that is always true for force symmetrisation by kernel averaging and true for symmetrisation by arithmetic averaging only for $h_{hot} >> h_{clump}$.  For harmonic averaging of the smoothing lengths, this reduction in drag as density increases is only partially effective for small $h_{hot}$ and disappears for $h_{hot} >> h_{clump}$.  Geometric averaging leads to the different relationship of $F_{drag} \propto - \rho_{hot}^{\twothirds}$, which is closer to the form normally expected.

To verify that this offsetting phenomenon occurs, a set of tests similar to those in \sect{sec.Results.Drag.h_ij_Scaling} were performed but with the mass resolution held constant while the number density is varied.  Hence, the clump was ploughed through material of both different physical densities as well as number density.  Since the velocity of sound varies only with the temperature of the gas, varying the mass density does not change the Mach number.

The measured deceleration for the clump in each test was then fit to the analytic form given by \eq{eq.Expected} in the same fashion as described in \sect{sec.Results.Drag.h_ij_Scaling}.  This analysis produced an effective radius which could then be compared with the smoothing radii of the clump and the hot particles as well as the averaged value. If the effective radius still scaled with the averaged smoothing radius, $h_{ij}$, then the offsetting phenomenon would be verified.  The tests and analysis were done for hot media with physical densities of 1, 2, and 5 times that employed in \sect{sec.Results.Drag.h_ij_Scaling} for the velocities corresponding to Mach~\onethird, Mach~1, and Mach~2.

The results of the analysis, summarised in \tab{tab.Results.VaryingRho}, confirm that as the density contrast between the clump and the background density decreases, the effective radius still scales with $h_{ij}$ and, as a consequence, the force of drag scales with $\rho_{hot}^{\onethird}$. In \tab{tab.Results.VaryingRho}, the scaling of the cross-section with the averaged smoothing radius, $h_{ij}$, (fifth column) is very clear and distinguishable from scaling with either of the other two smoothing radii, $h_{clump}$ or $h_{hot}$ (third and fourth columns).  This is true independent of the velocity regime.

\subsubsection{The method of force symmetrisation}
\label{sec.Results.Drag.symm_method}
Since the effective radius of the clump scales with the averaged smoothing length, $h_{ij}$, then it is to be expected that the method of force symmetrisation, which requires the use of an averaged smoothing length, should affect the drag. Verification of this is significant, since the method of force symmetrisation is something that is chosen during the coding of SPH.

For the method of force symmetrisation provided by using a common value, $h_{ij}$, of the smoothing lengths calculated by arithmetic, harmonic, or geometric averaging, it is easy to show that arithmetic averaging will always produce the largest value of $h_{ij}$ while harmonic averaging leads to the smallest. Averaging the kernels themselves does not lend itself to such a comparison, since the effective value of $h_{ij}$ would be dependent on $r$.  However, using the argument to estimate $R_{eff}\simeq 2h_{ij}$ outlined in \sect{sec.Expectation}, for kernel averaging we have $h_{ij} \simeq h_{hot}$, which is a maximum. \Fig{fig.SPH.KernelShapes} verifies that at large radii, averaging the kernels indeed leads to the larger weighting and, hence, degree of interaction.

For those model systems in which the ratio of $h_{hot}$ to $h_{clump}$ is $2.7$ times, the differences between the method of force symmetrisation should be quite appreciable; averaging the kernel produces an estimated radius of interaction almost twice that of harmonic averaging of the smoothing radii.  This is consistent with the results of \citet{Thacker98} in trend, if not magnitude, in which it is noted that averaging the kernels leads to the largest amounts of drag at velocities less than Mach~1 while harmonic averaging of the smoothing lengths leads to the least.

As a test, the drag on a constant-velocity Mach~1 clump was measured while varying the method of symmetrisation.  The implementations used kernel, arithmetic and harmonic averaging schemes.  Drags of $1.4\expd{36}$, $1.0\expd{36}$, and $0.6\expd{36} \dyne$ were measured respectively. The test was not performed for geometric averaging of the smoothing lengths since this method produces a value for $h_{ij}$ intermediate between arithmetic and harmonic averaging.  These results confirm the trend that larger values for $h_{ij}$ produce a greater drag, with the kernel averaging scheme producing the largest mean drag, 2.4 times larger than that produced with harmonic averaging of the smoothing lengths and 1.4 times larger than that found when arithmetic averaging is implemented.  However, since the drag should scale with $R_{eff}^2$, kernel averaging should produce four times as much drag, instead of the 2.4 measured.  This indicates that there are secondary effects produced by the method of symmetrisation that, for these tests, offset the direct connection between $h_{ij}$ and drag.  However, it is clear that the choice of method for force symmetrisation significantly affects the amount of drag incurred.

\subsubsection{Energy deposition}
\label{sec.Results.Drag.Energy}
\begin{figure}
\epsscale{\figscaletwo}
\plotone{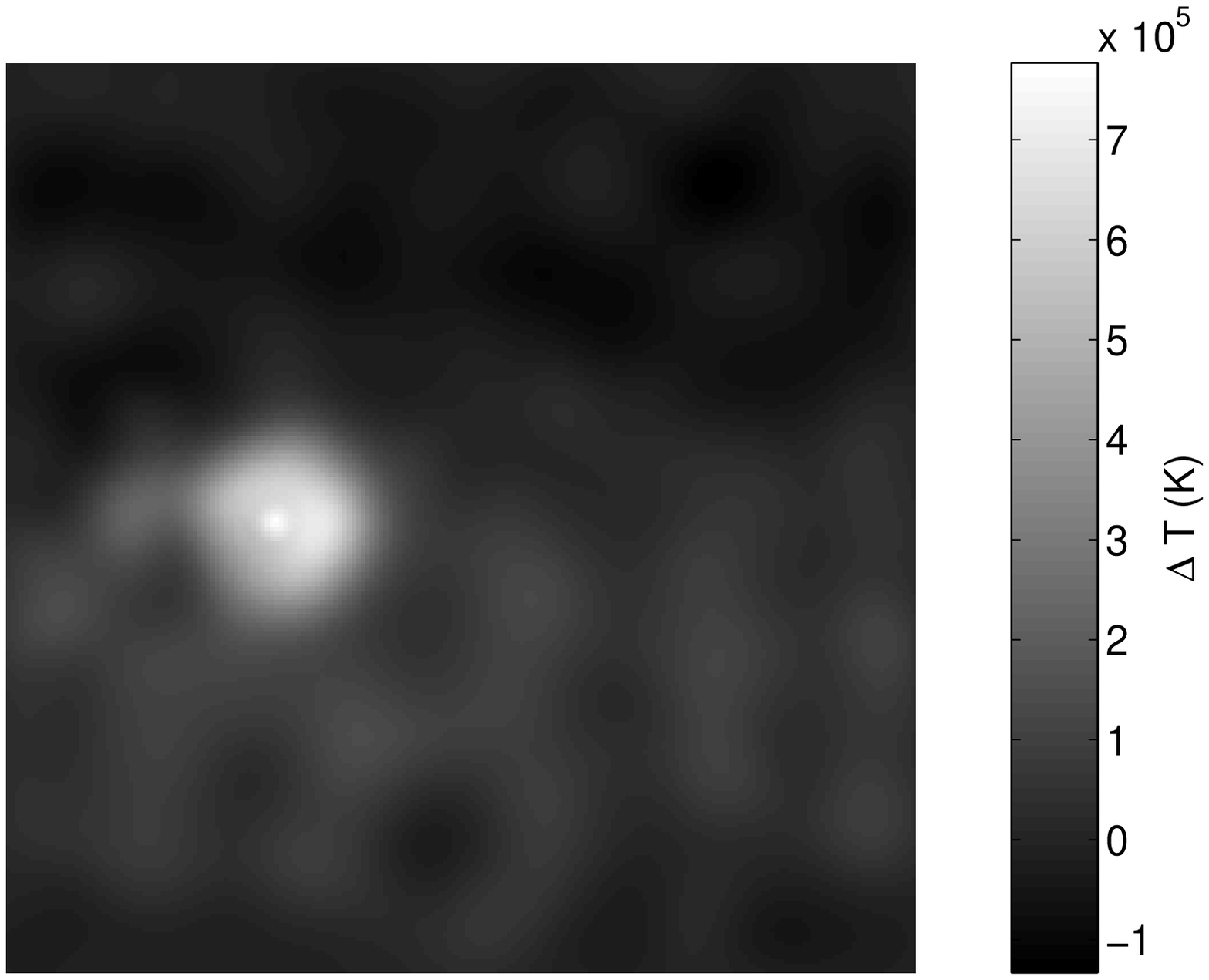} \\
\plotone{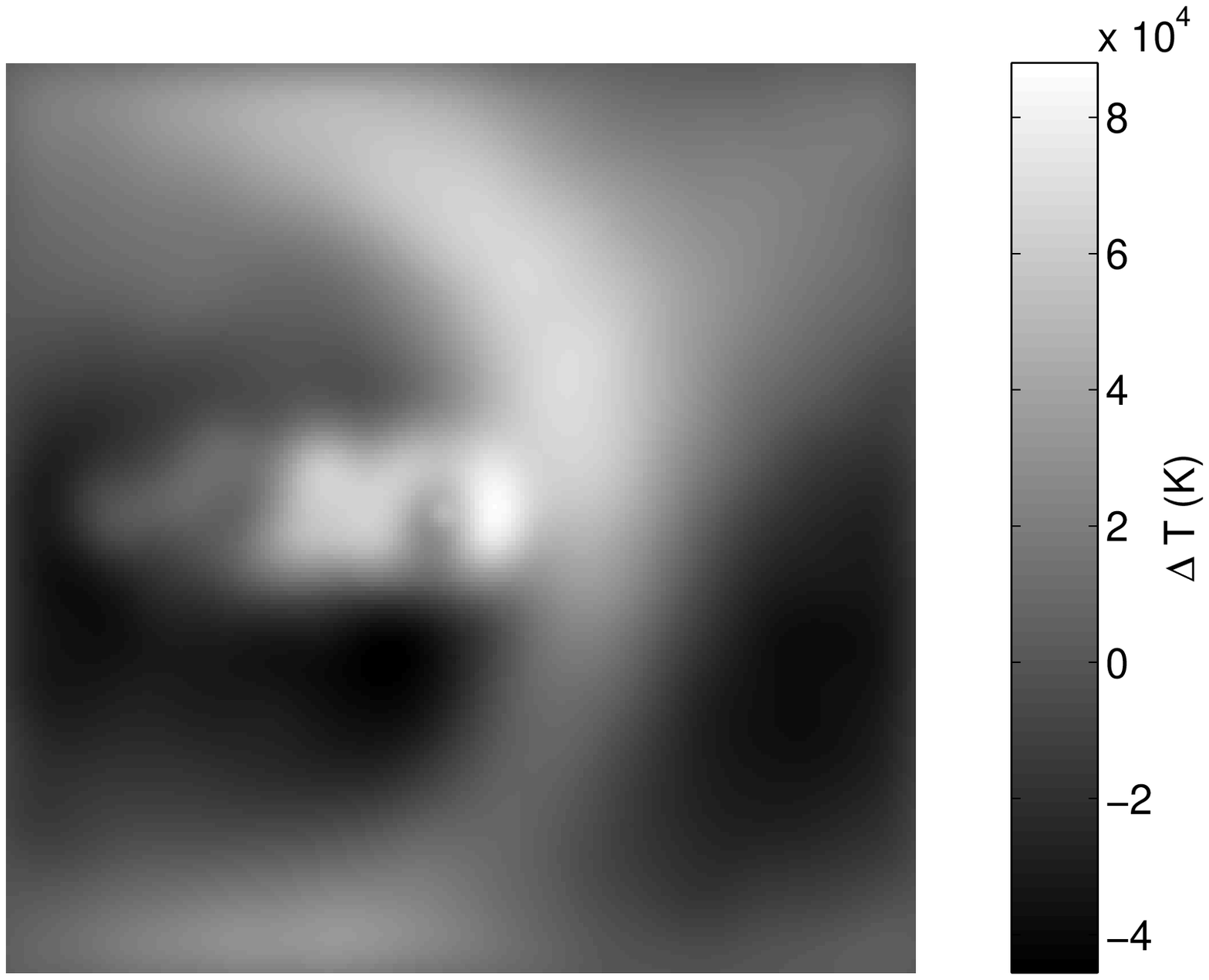} 
\plotone{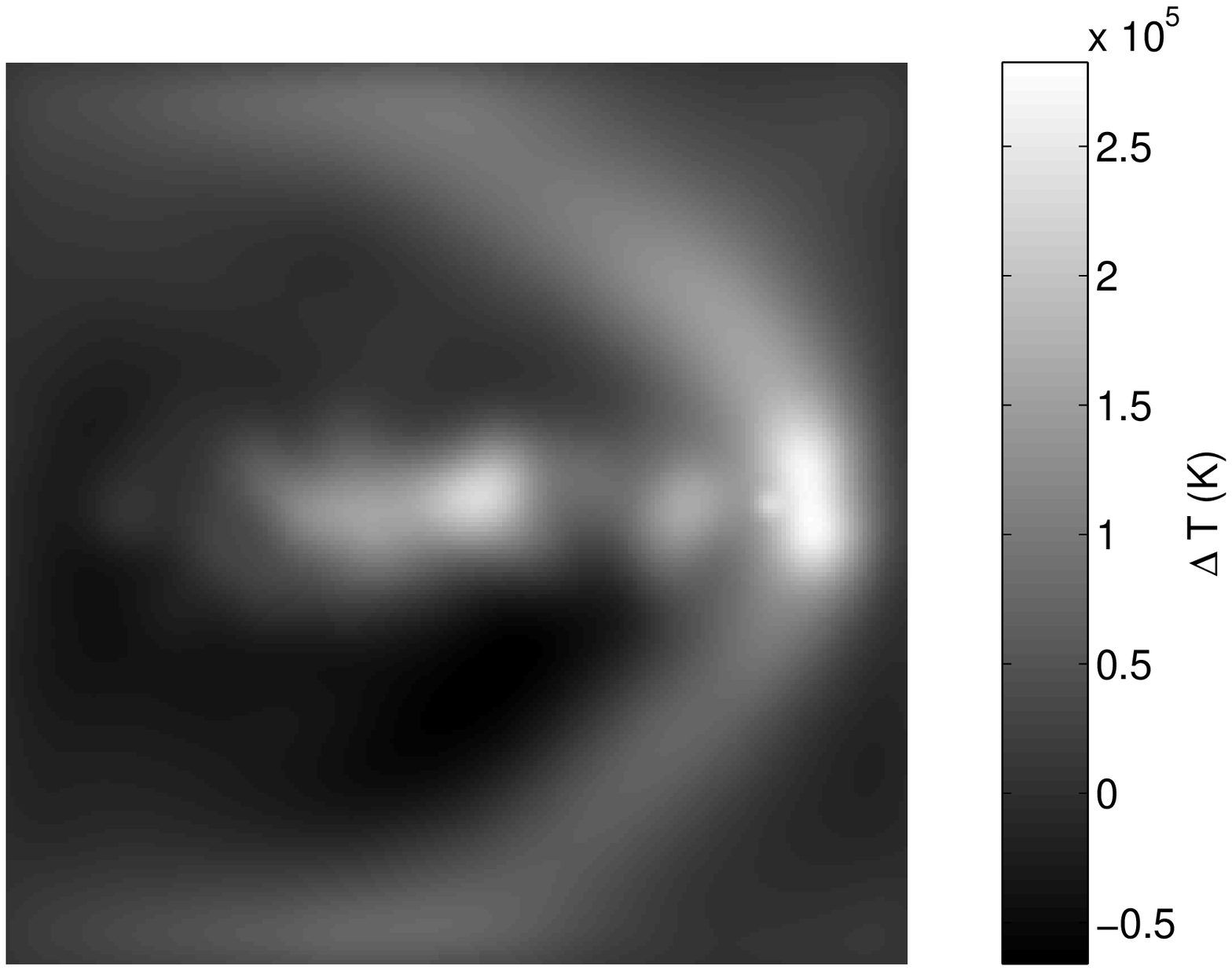}
\caption{The temperature change in the hot gas due to the passage of the clump.
The temperature changes for the gas is plotted for the Mach \onethird\ (top), Mach 1 (left), and Mach 2 clumps (right). White indicates a larger temperature increase, black indicates no change.  Note that the temperature changes are small compared to the actual temperature of the gas ($10^5 \K$ cf. $10^7 \K$).  The clump starts at a position one tenth of the box size from the left. The slices are $10 h_{clump}$ thick.
}
\label{fig.Results.dT}
\end{figure}

The energy lost by the deceleration of the clump goes primarily into heating the surrounding gas.  Since this gas is already hot, this extra thermal energy is absorbed without any significant change in the temperature of the gas.  The total energy of the clump is less than $1\%$ of the total thermal energy of the hot gas.  The spatial distribution of this deposition of energy into the hot gas can be examined by looking at the {\em change} in temperature of the gas
particles.  This is illustrated in \fig{fig.Results.dT}. As expected, a bow shock is formed in the sonic and supersonic cases. On account of the large filling factor of the shock, it contains a large fraction of the deposited energy.  Energy is also deposited in the wake of the clump.

For the subsonic clump, a transitory pulse is created at the start of the simulation which quickly dissipates.  Subsequently, most of the energy is deposited locally and carried along with the clump.

\subsection{Drag in the subsonic regime}
\label{sec.Results.Subsonic}
It was established in \sect{sec.Results.Drag.Excess} that subsonic clumps have inappropriately large effective cross sections. This is demonstrated by \tabs{tab.Results.VaryingNden}~and~\ref{tab.Results.VaryingRho} and \figs{fig.Results.ReffVsMach}~and~\ref{fig.Results.MeanClumpDrag}.  Whereas the sonic and supersonic clumps always decelerate less than the fiducial rate, the subsonic clumps experience a greater drag than expected.  We show in this section that the clump accretes a layer of hot gas particles into a shell at the effective smoothing radius, $2h_{ij}$, and it is this shell that enhances the drag.  Furthermore, it is demonstrated that this is a by-product of the SPH pressure calculation in the vicinity of a compact collection of cold particles coupled to the artificial viscosity which damps rapid accelerations.  In this configuration, the calculated force on a hot gas particle due to gas pressure causes oscillations in the hot particle's velocity which triggers the viscosity term of the hydrodynamic forces, bringing the hot particle to rest relative to the moving clump.  In the subsonic regime we expect the clump to accrete some material as it moves through the hot medium, but it is the increase in cross section due to this shell of accreted particles that causes the clump to slow excessively.

\subsubsection{The accretion of remoras}
\label{sec.Results.Subsonic.remoras}
Using the data from \sect{sec.Results.Drag.h_ij_Scaling} in which a cold clump was allowed to decelerate in a uniformly dense medium of hot gas, the behaviour of the drag on a subsonic clump may be characterised. The subsonic clump initially follows approximately the same deceleration curve as given by \eq{eq.Expected}.  The transition to a more rapid deceleration occurs as particles are accreted from the hot gas phase.  Accretion of these particles is evident in \fig{fig.Results.flow2} which displays the flow of particles about the cold clump in the reference frame of the cold clump at various Mach numbers. No hot particles are decoupled from the flow of hot particles in the Mach~1 and Mach~2 simulations while 44 are decoupled in the Mach~\onethird\ run.

The particles picked up from the hot medium will be referred to as `remoras'. Since remoras are also found about the subsonic clumps which are not permitted to decelerate (as in \sect{sec.Results.Drag.Excess}), it is clear that the accretion does not occur only as the clump slows to almost stationary, but occurs while the clump still has a significant velocity.

The remoras form a narrow shell about the clump at a radius equal to the effective smoothing radius, $2h_{ij}$.  Note that in this section we are using simple arithmetic averaging for the combined smoothing length; that is $h_{ij}=(h_{hot} + h_{clump})/2$.  This shell is indicated by \fig{fig.Results.flow2} and is verified in \fig{fig.Results.CorrFunction_2001}, which plots the 2-point correlation function about the clump particles.  The correlation function of the hot gas particles (dashed line in \fig{fig.Results.CorrFunction_2001}) shows that the particles have preferential spacings typical of a glass distribution, but not at the distance of $2h_{ij}$.  This association with the radius $h_{ij}$ is strong evidence that the SPH calculations are at the root of this phenomenon.

The tight span of radii in which the remoras are permitted to come to rest is demonstrated by the narrowness of the peak in the correlation function at $h_{ij}$, witnessed in \fig{fig.Results.CorrFunction_2001}.  The jump in the correlation function is two orders of magnitude and is located just beyond $2h_{ij}$ with a width of $15\kpc$ or about $10\%$ of $2h_{ij}$.

\subsubsection{SPH forces for a particle near $2 h_{ij}$}
\label{sec.Results.subsonic.SPH}
The presence of this shell of remoras about the cold clump is clearly a by-product of the hydrodynamical calculations; it is not seen in collisionless simulations and the radius of the shell is coincident with the averaged smoothing radius.  Here, we will examine the behaviour of the SPH force calculation for a hot gas particle in the vicinity of a dense knot of cold gas.  It will be shown to lead to a process that traps the hot particles at the radius, $2h{ij}$, seen for the remoras in the simulations.

In SPH, the pressure forces on an individual particle, $F_i$, are summed symmetrically over all neighbouring particle pairs in the manner of \eq{eq.tests.intro.ForceCalcSM93} which can be rewritten
\begin{equation}
F_i  \sim \sum_j^{N} -m_i m_j \left( \frac{T_i}{\rho_i} + \frac{T_j}{\rho_j} \right) \grad W(r_i-r_j,h_{ij}).
\label{eq.Results.SPHpressure} 
\end{equation}
This utilises SPH's ability to estimate gradients of local scalar quantities by scaling the local values of these quantities (the pressure, $P \sim \rho T$, here) by the gradient of the kernel function, $W$. This is normally very effective when the local gradients in the temperature and, more significantly, the gas density, $\rho$, are small.  While the temperature is a parameter accumulated over time, $\rho$ is a quantity re-estimated each timestep.  Hence, in the presence of the steep density gradients surrounding cold clumps, this value can fluctuate significantly between iterations.  If a hot gas particle is within twice its smoothing length, $2h_{hot}$, of the centrally-concentrated cold clump, then the estimated gas density will be significantly larger ($1.5\times$, here) for the particle than otherwise.

It is informative to determine what the summation of the pressure forces is for a hot gas particle between $2h_{ij}$ and $2h_{hot}$ from a cold clump.  There will be two regimes: one in which the gas particle is at the far end of this distance, and the other when it is near $2h_{ij}$.  In the first case, the hot gas particle will not be surrounded by an isotropic distribution of particles.  There will be a void on the side of the particle facing the cold clump.  The 2-point correlation function of the particles in the vicinity of the cold clump (\fig{fig.Results.CorrFunction_2001}) demonstrates the existence of this void which occurs because particles are repelled by pressure forces close to the cold clump. The particles in the cold clump are centrally condensed within a radius much smaller than their SPH smoothing radius which is, in turn, half the gravitational softening radius (the minimum value we allow it to take).  Surrounding this clump is a void with a radius slightly smaller than the smoothing radius used in the SPH calculations when dealing with a pair of hot and cold particles.  Hence, the average gradient will force the particle toward the cold clump.

Once the particle is within $2h_{ij}$ of the cold clump it suddenly sees many more than $N_{SPH}$ neighbours so its smoothing length drops while its density estimate jumps. For a very diffuse hot medium essentially all the neighbours of the particle are now in the cold clump and the particle is repelled strongly by pressure forces.  The kick the particle receives from the interaction with the cold clump is short lived, and soon the unbalanced pressure force from the surrounding hot gas forces the particle back towards the clump starting the cycle again.

A hot gas particle near a cold clump thus oscillates (or perhaps, more correctly, bounces) just beyond the radius, $2h_{ij}$.  This oscillation is a consequence of the form of the SPH force calculations coupled with its treatment of neighbours and, as such, will be a characteristic of any implementation of SPH.

The path of a particle that is just grazed by the cold clump as it passes is shown in \fig{fig.Results.Flow_1439}. The path, in the reference frame of the cold clump, shows the series of accelerations described. What is even clearer is how these accelerations conspire to trap the particle at a radius approximately half way between the smoothing radii for the clump and the hot gas.

In \fig{fig.Results.ZForces_1439}, which plots the hydrodynamic forces for the hot gas particle in a direction perpendicular to the direction of motion of the clump, the alternating directions of the forces are clearly illustrated. In this figure, the density of the hot particle, as estimated by the SPH density calculation, is shown at the bottom.  It clearly delineates the two regimes described previously; when the density is low, the particle does not `see' the clump whereas when it is high, it is under the influence of the clump and feels a strong force away from it.

The oscillations between these two regimes are important as they trigger the artificial viscosity which drags the particle into moving with the clump.  As the viscosity is only triggered if two particles are converging, the particle is interacting viscously alternately with the clump and the hot gas.  The viscosity attempts to damp the oscillations, which can also be seen in \fig{fig.Results.ZForces_1439}. The attempt to damp the oscillations has the by-product of turning the particle into a remora; as \fig{fig.Results.CumulativeMom_1439} shows, though the pressure is still the dominant force, the accumulated momentum in the direction of travel is controlled by the contributions from the viscosity term.

\subsubsection{The significance of a compact core}
\label{sec.Results.Subsonic.Antigrav}
The previous description of the problem indicates that the compact nature of the clump exacerbates the problem of the creation of a shell of remoras; the hot particles are influenced by either all or none of the clump particles causing large swings in the pressure forces.  The effect may be mitigated if the hot particles come under the influence of the cold particles gradually, instead of abruptly as they approach the clump.  This may be tested by having the clump particles occupy a larger volume within their smoothing radius.  Presented here is the method by which this was done and the resulting effect on the drag.

The cold clump particles may be forced to form a more extended clump by modifying the form of the gravitational potential to produce a repulsive force at close distances.  The unmodified gravitational force expression is a spline approximation of a Plummer-softened $1/r$ potential.  This was replaced by a gravitational potential of the form
\begin{equation}
U_g = \frac{1-(x+a)^{-n}}{x+1}
\label{eq.Results.AntiGravPot}
\end{equation}
where $x=r/\epsilon_g$ with $\epsilon_g$ being the gravitational softening parameter. Experimentation found that the parameters $a=0.2$ and $n=2$ suitably dispersed the clump, increasing the mean interparticle spacing by a factor of 100, but keeping the clump particles within twice the gravitational softening length.

The influence of the antigravity factor extends to beyond twice the gravitational softening length to compensate for the strong compressional forces exerted by the pressure of the surrounding hot gas.  As such, this form does not permit a stable clump to exist in isolation.  This also does not permit a simple drag experiment; by the time the clump has expanded to a new stable configuration within the hot gas, the clump has already decelerated appreciably. Instead we again hold the velocity of the cold clump fixed as it progresses
through the hot medium.  This allows a stable configuration to develop and the drag on the clump to be measured.

The drag for the compact clump was measured to be $6.0\expd{35} \dyne$ while the drag on the more diffuse clump was $4.2\expd{35} \dyne$.  The diffuse clump experiences almost \onethird\ less drag.

This factor is remarkable, considering that the diffuse clump has a larger cross-section.  The cross-sections may be measured by convolving the clump particle positions with their effective smoothing radius for interaction with the hot particles, $h_{ij}$, to produce a column density distribution.  Since the hot gas particles are influenced strongly at even small column densities, we take $10\%$ of the maximum in the column density distribution for the compact core to calculate the total effective cross section of the clumps.  The extended clump is found to have a cross-section that is $60\%$ larger than that of the compact core.

Taken together, that the diffuse clump experiences \onethird\ less drag while having a cross section $60\%$ larger implies that the effect of the abrupt transition between the cold clump and the hot particles caused by having a very small central clump doubles the drag.  This is a clear demonstration of the significance of a compact core in enhancing the drag.

\citet{Thacker98} notes in a similar set of tests a difference in the clump sizes depending on the viscosity form adapted.  It reports that a Monaghan-type artificial viscosity, as used in these tests, produces a clump that is two to three times more extended that those that use a $\divdotv$ form.  It should be emphasized that the mean interparticle spacing of the cold clump particles was required to be expanded by a factor of one hundred before the drag was diminished. Hence, the increase in drag is not something that can be easily fixed by modifying the implementation of SPH.  This is particularly true since expanding the clump leaves it more susceptible to disruption through pressure and tidal stripping, a feature that would worsen the overmerging problem.

\section{DISCUSSION}
\label{sec.Discussion}
Using a series of simulations of a cold clump moving through a hot medium, we have explored the nature of drag in scenarios directly relevant to cosmological structure formation.  The results are scale invariant and, subsequently, may be applied to astrophysical situations non-cosmological in nature.  Indeed, any hydrodynamical simulation using the SPH technique in which drag on resolution-limited objects is significant will be affected by the results described herewith.

An expression for a fiducial drag of the clump was determined analytically presuming completely inelastic collisions.  The expression derived, when supplemented by a coefficient of drag, is identical to previously derived analytic forms.  The force goes as $F_{drag} \propto V^2 R^2$.  An estimate was made for the cross-sectional radius of the clump which was found to be twice the averaged smoothing length, $h_{ij}$.  This is an average of the SPH smoothing length common to the cold clump particles, $h_{clump}$, and that common to the hot gas particles, $h_{hot}$, which is used to symmetrise the SPH force calculations.  In the case of symmetrisation via kernel averaging, the radius of the clump should approximate $2 h_{hot}$.

The energy lost from the simulated clumps as they decelerate is deposited locally in the case of the subsonic clump. For a sonic and supersonic clump, SPH successfully models a bow shock which carries away the energy from the clump, depositing it over a larger volume.

The drag was compared with the fiducial amount by equating the measured drag to the fiducial form and solving to determine an effective radius, $R_{eff}$, for the clump.  Over the entire range examined, the effective radius is fit quite well by the form, $R_{eff} \propto \Mach^{\fraction{-3}{5}}$.  This implies that $F_{drag} \propto \Mach^{\fraction{4}{5}}$, which is almost linear.

In the regime $\Mach > \threefourths$, the effective radii of the simulated clumps lie in the range $R_{eff} \simeq h_{ij}$ to $2 h_{ij}$.  This is consistent with the estimated value for the effective radius of $2 h_{ij}$, particularly if the coefficient of drag is between $0.5$ and 1.  However, there is no evidence of convergence to any particular value of $R_{eff}$ as $\Mach$ increases.

For the range $\Mach < \threefourths$, the effective radius is larger than the estimate; i.e., $R_{eff} > 2 h_{ij}$.  Indeed, it is larger than even the maximum radius of direct hydrodynamic influence, $2 h_{hot}$.  This demonstrates that the drag on sub-sonic clumps exceeds any physically justifiable amount presuming inelastic collisions.

It is notable that the transition point, $\Mach \simeq \threefourths$, is independent of the local velocity of sound.  This demonstrates that the simulations are velocity-scale invariant.

At all velocities, the drag force was found to scale with the cross-sectional area provided by the averaged smoothing length, $h_{ij}$. There are two immediate implications of this: the method of force symmetrisation (which determines the method of averaging) affects the drag, and the effective cross section of the clump varies with the density of the medium through which it is travelling.

In regards to the first implication, the method of force symmetrisation is selected during the coding of the SPH implementation. It can be one of the following four methods: arithmetically averaging the kernels, or averaging the smoothing radii of the interacting pair of particles using either arithmetic, harmonic, or kernel averaging.  It was verified that harmonic averaging of the smoothing radii produces the smallest effective cross-section whereas kernel averaging produces the largest.  The effect is substantial, with the drag produced by the kernel averaging $2.4$ times larger than that of the harmonic averaging.  The drag forces from geometric and arithmetic averaging of the smoothing lengths fall in between, with arithmetic averaging the larger of the two.

Concerning the second implication of the cross-sectional radius scaling with $h_{ij}$, it was shown that there is a loss of proper scaling of the drag with density.  The drag should scale with density as $F\propto \rho$.  This occurs because the average smoothing length, $h_{ij}$, varies with the density of the surrounding medium.   However, higher densities lead to a smaller average smoothing length, which decreases the cross-sectional area of the clump, offsetting the increase in drag due to the increased density.  This is particularly true for methods of force symmetrisation that use kernel averaging or arithmetic averaging of the smoothing lengths.  For these, the drag on a clump is proportional to $\rho_{hot}^{\onethird}$ instead of $\rho_{hot}$.  For geometric averaging, this relation changes to $\rho_{hot}^{\twothirds}$, while for harmonic averaging, the effect is present over a limited range of $h_{hot}/h_{clump}$.

This attenuation of the drag as the density increases will have consequences in simulations of the hierarchical formation of cosmological structure. A disproportionate amount of deceleration will occur in the outer halos of the structures.  This will preferentially deposit energy in the outer halo as well as alter the spatial distribution of the deposited matter.  This should directly affect the velocity and spatial bias of the baryonic material in the structures.

More disturbing is the dramatically large drag felt by subsonic clumps.  This was demonstrated to be a by-product of the method of force calculation used by SPH. For clumps travelling at $\Mach < \threefourths$, the method SPH uses to calculate the hydrodynamical forces, coupled with the compact nature of the cold clump, initiates a process that accretes hot particles onto a sphere about the cold clump. The hot gas particles are forced by the interaction with the cold clump to oscillate outside the radius provided by twice the averaged smoothing radius, triggering the artificial viscosity which boosts the velocity of the hot particles to match that of the cold clump. This process is inherent to the SPH method since it arises from the force calculation.  It leads to the excessive drag at low velocities because the particles accreted by the cold clump, or remoras, increase the effective cross-sectional area of the clump.

\begin{figure}
\epsscale{\figscaletwo}
\plotone{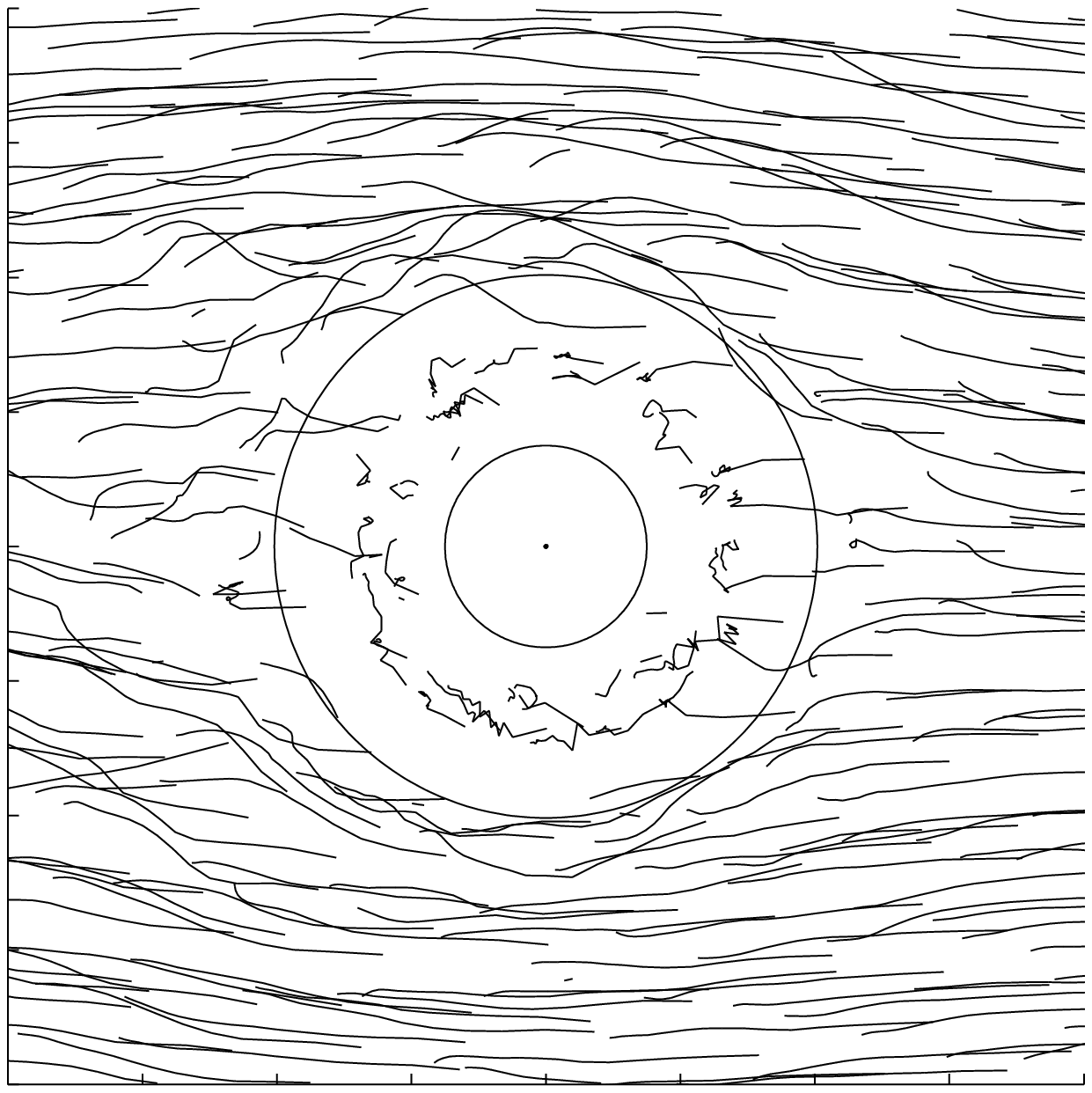} \\
\plotone{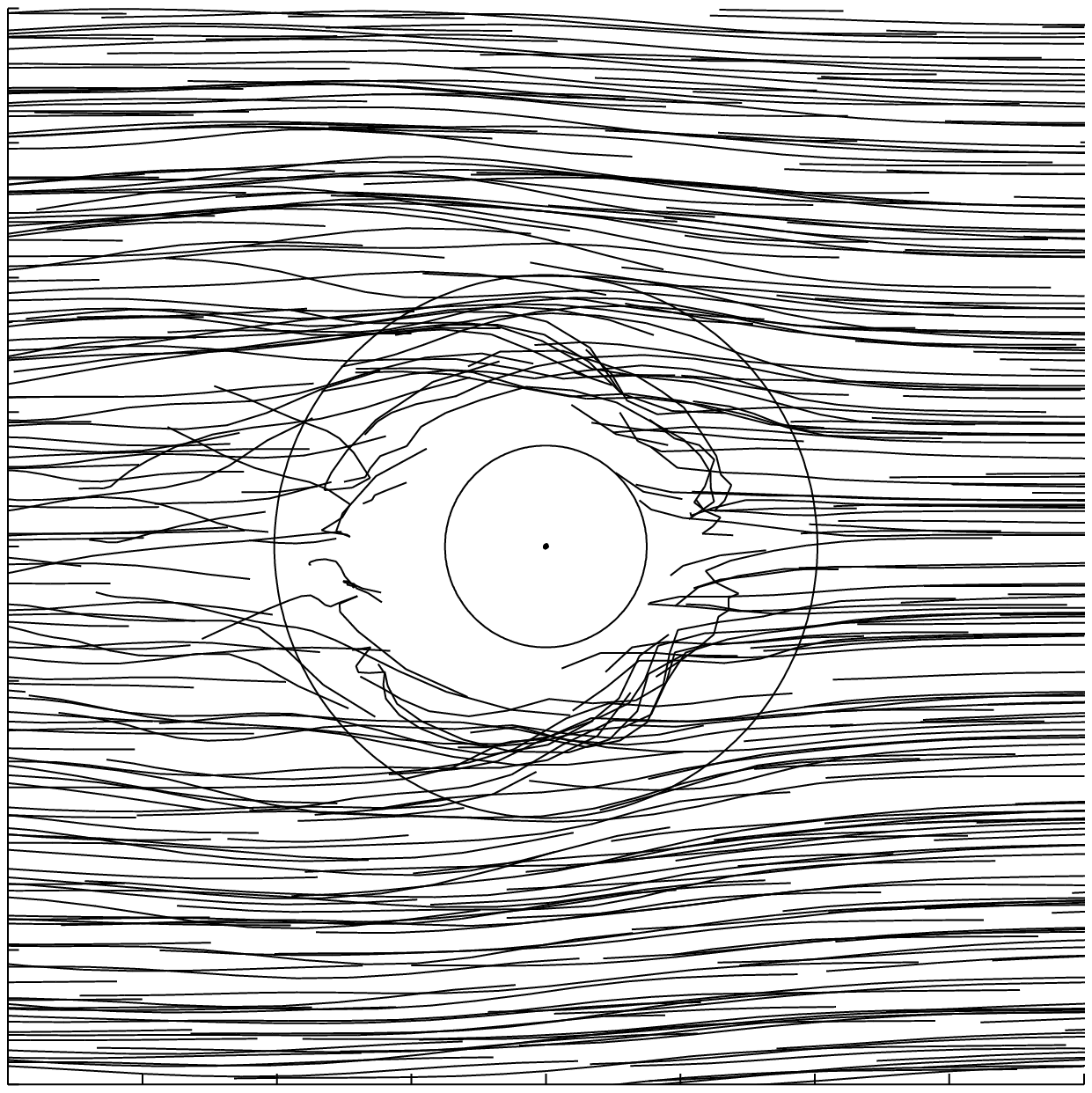}
\plotone{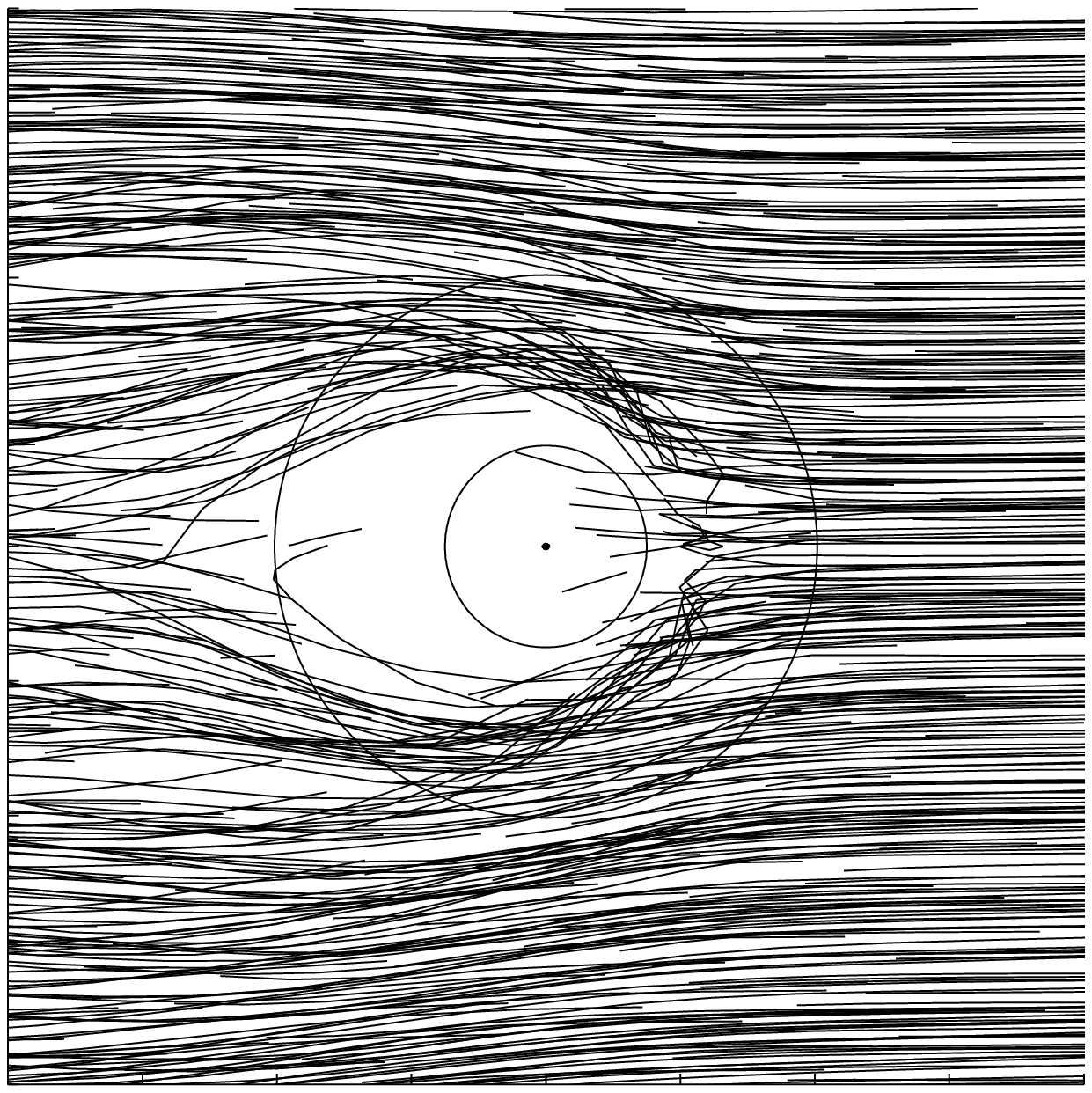}
\caption{Flow of halo particles about clumps.  The flows are plotted
for the Mach \onethird\ (top), Mach 1 (left), and Mach 2
clumps (right). The inner circle is the radius $2h_{clump}$ while the outer circle is the same for the hot gas, $2h_{hot}$. The effective smoothing length, $h_{ij}$, for this simulation is simply the arithmetic mean of $h_i$ and $h_j$. Flow is from right to left.  The slices are $5h_{clump}$ thick.}
\label{fig.Results.flow2}
\end{figure}

The accretion of the remoras leads to an increase in drag by increasing the effective cross section. The flow of the particles in \fig{fig.Results.flow2} indicates that the gas particles are not being repulsed solely by the clump, but also by the remoras in the shell about the clump.  If this is the case, then the cross sectional radius should increase from $2h_{ij}$ to $2(h_{ij}+h_{hot})$.  For the systems presented here, this should increase the cross-sectional radius by $2.5$ times.

\begin{figure}
\epsscale{\figscale}
\plotone{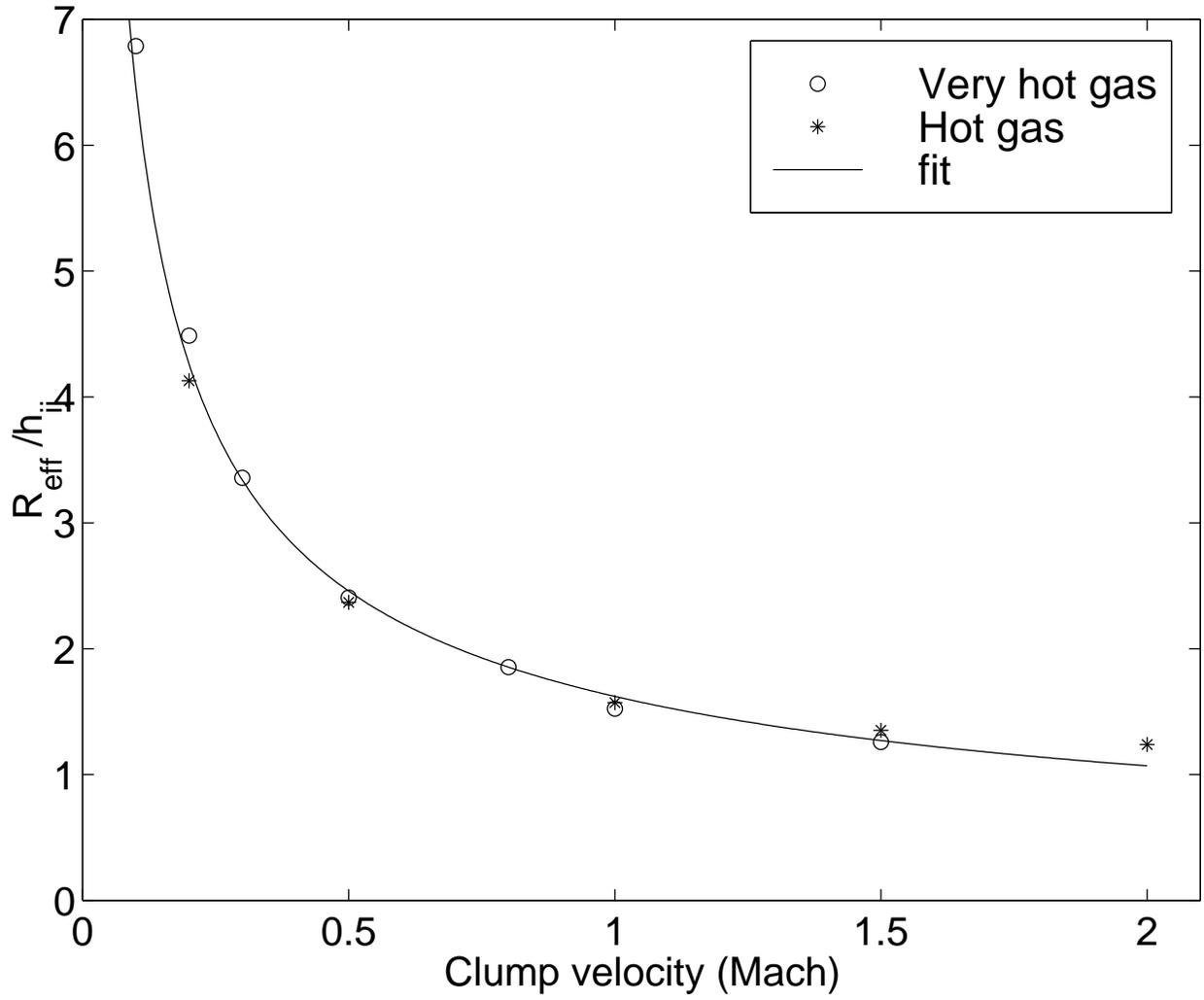}
\caption{The effective radius of the clump as it varies with Mach number.
The data are given for both the very hot gas ($V_s = 1500 \kps$) and
the hot gas ($V_s = 500 \kps$).}
\label{fig.Results.ReffVsMach}
\end{figure}

Comparing the effective radii calculated for the Mach~\onethird\ clump with those of the Mach~1, and Mach~2 clumps imply respective ratios of $2.3\pm 0.3$ and $2.9\pm 0.4$ (see \tab{tab.Results.VaryingNden} for the radii).  However, \fig{fig.Results.ReffVsMach} shows that the change in $R_{eff}$ does not occur as a step function, nor does it plateau at $2.5 h_{ij}$.  This indicates that further processes are occurring. One possibility is `crystallisation' of the clump into the matrix of the hot gas particles as the velocity of the clump approaches the velocity dispersion of the clump \citep{LSS98}.

For virialised objects, the velocity dispersion of the component structures should be on the order of the speed of sound of the halo gas.  This is approaching the velocity at which the accretion of remoras commences and we observe an inappropriate increase in the drag.  This will enhance the merging rate of the structures. Not only will this modify the distribution of small-scale structures, but it, too, will affect the velocity and spatial bias of the baryons.

Boosting the volume in which the cold clumps reside by providing a short-range anti-gravity force is shown to reduce the drag by a factor of two.  This is not a viable solution to the problem, however, since the clumps are then not stable outside a hot halo.  What may be required is a decoupling of the separate hot and cold phases, similar to that described in \citet{Pearce99}, with the inclusion of a solid-body force between the clump and the hot gas.

\acknowledgements 
We acknowledge NATO CRG 970081 which facilitated our interaction.
We would like to thank Dr. R. Thacker for supplying
the various implementations of SPH. This work was supported 
by NSERC Canada.

\bibliographystyle{plainnat_mnras}
\bibliography{biblio}

\renewcommand{\thisdir}{Tables/tables}
\begin{table}
\begin{center}
\begin{tabular}{cccc}
                  & Slow cold clump & Fast cold clump \\
\hline
$\rho/\rho_c$     & $1000$          & $1000$          \\
$T$ ($\K$)        & $10^4$          & $10^4$          \\
$R$ ($\kpc$)      & $50$            & $50$            \\
$N$               & $100$           & $100$           \\
$m$ ($10^9\Msun$) & $1.7$           & $1.7$           \\
$V_o$ ($\kps$)    & $500$           & $1000$          \\
\hline
\hline
                  & Hot gas         & Very hot gas    \\
\hline
$\rho/\rho_c$     & $10$            & $10$	      \\
$T$ ($\K$)        & $10^7$          & $10^8$          \\
$N$               & $13000$         & $13000$         \\
$m$ ($10^9\Msun$) & $1.7$           & $1.7$	      \\
$V_s$ ($\kps$)    & $500$           & $1500$          \\
$R_J$ ($\Mpc$)    & $6$             & $18$	      \\
\hline
\end{tabular}
\end{center}
\caption[The characteristics of the drag tests]
{The characteristics of the cold clumps and the hot media used in the drag tests. Given are the overdensity, $\rho/\rho_c$ ($h_{100}=1$), the temperature, $T$, the radius of the cold clump, $R$, the number of particles in the medium, $N$, the mass resolution of the medium, $m$, the initial velocity of the cold clump, $V_o$, the speed of sound in the hot medium, $V_s$, and the Jeans length for the hot medium, $R_J$. The simulation volume in all cases is $(2 \Mpc)^3$.  The `fast cold clump' was used in the Mach 2 runs in combination with the `hot gas'.  The Mach 1 runs used the `slow cold clump' embedded in the `hot gas'.  The Mach 1/3 runs used the `slow cold clump' in the `very hot gas'.}
\label{Tab.Drag.Init}
\end{table}

\begin{table}
\begin{center}
\begin{tabular}{ccccc}
\multicolumn{5}{c}{Mach \onethird} \\
\hline
$n_{hot}$      & $R_{eff} (\Mpc)$ & $R_{eff}/h_{clump}$ & $R_{eff}/h_{hot}$ & $R_{eff}/h_{ij}$ \\
$\times 1$ & 0.23  &  6.0  &  2.2  &  3.3 \\
$\times 2$ & 0.17  &  4.4  &  2.1  &  2.8 \\
$\times 5$ & 0.15  &  4.0  &  2.6  &  3.1 \\
\hline\hline
\multicolumn{5}{c}{Mach 1} \\
\hline
$n_{hot}$      & $R_{eff} (\Mpc)$ & $R_{eff}/h_{clump}$ & $R_{eff}/h_{hot}$ & $R_{eff}/h_{ij}$ \\
$\times 1$ & 0.13  &  3.6  &  1.3  &  1.9 \\
$\times 2$ & 0.11  &  3.0  &  1.4  &  1.9 \\
$\times 5$ & 0.07  &  1.9  &  1.2  &  1.5 \\
\hline\hline
\multicolumn{5}{c}{Mach 2} \\
\hline
$n_{hot}$      & $R_{eff} (\Mpc)$ & $R_{eff}/h_{clump}$ & $R_{eff}/h_{hot}$ & $R_{eff}/h_{ij}$ \\
$\times 1$ & 0.093  &  2.5  &  0.9  &  1.3 \\
$\times 2$ & 0.086  &  2.3  &  1.1  &  1.5 \\
$\times 5$ & 0.063  &  1.7  &  1.1  &  1.3 \\
\hline\hline
\end{tabular}
\end{center}
\caption{Scaling of the effective radius, $R_{eff}$, of the clump with mean smoothing length of the clump gas particles, $h_{clump}$, the hot gas, $h_{hot}$, and the symmetrised smoothing length, $h_{ij}$, as used in the SPH calculations.  Varied is the number density of the hot gas particles, $n_{hot}$, but not the mass density.}
\label{tab.Results.VaryingNden}
\end{table}

\begin{table}
\begin{center}
\begin{tabular}{ccccc}
\multicolumn{5}{c}{Mach \onethird} \\
\hline
$\rho_{hot}$      & $R_{eff} (\Mpc)$ & $R_{eff}/h_{clump}$ & $R_{eff}/h_{hot}$ & $R_{eff}/h_{ij}$ \\
$\times 1$ & 0.25 & 6.7 & 2.5 & 3.6 \\
$\times 2$ & 0.22 & 5.9 & 2.8 & 3.8 \\
$\times 5$ & 0.19 & 5.0 & 3.2 & 3.9 \\
\hline\hline
\multicolumn{5}{c}{Mach 1} \\
\hline
$\rho_{hot}$      & $R_{eff} (\Mpc)$ & $R_{eff}/h_{clump}$ & $R_{eff}/h_{hot}$ & $R_{eff}/h_{ij}$ \\
$\times 1$ & 0.13 & 3.6 & 1.3 & 1.9 \\
$\times 2$ & 0.12 & 3.1 & 1.4 & 2.0 \\
$\times 5$ & 0.09 & 2.4 & 1.5 & 1.9 \\
\hline\hline
\multicolumn{5}{c}{Mach 2} \\
\hline
$\rho_{hot}$      & $R_{eff} (\Mpc)$ & $R_{eff}/h_{clump}$ & $R_{eff}/h_{hot}$ & $R_{eff}/h_{ij}$ \\
$\times 1$ & 0.097 & 2.6 & 1.0 & 1.4 \\
$\times 2$ & 0.089 & 2.4 & 1.1 & 1.5 \\
$\times 5$ & 0.068 & 1.8 & 1.2 & 1.4 \\
\hline\hline
\end{tabular}
\end{center}
\caption{Scaling of the effective radius, $R_{eff}$, of the clump with
mean smoothing length of the clump gas particles, $h_{clump}$, the hot
gas, $h_{hot}$, and the symmetrised smoothing length, $h_{ij}$, as
used in the SPH calculations.  Varied is the physical density of the
hot gas particles, $\rho_{hot}$.}
\label{tab.Results.VaryingRho}
\end{table}

\clearpage

\renewcommand{\thisdir}{Figures/SPH}

\renewcommand{\thisdir}{Figures/Results}

\renewcommand{\figscale}{0.49}

\renewcommand{\figscale}{1}
\begin{figure}
\epsscale{\figscale}
\plotone{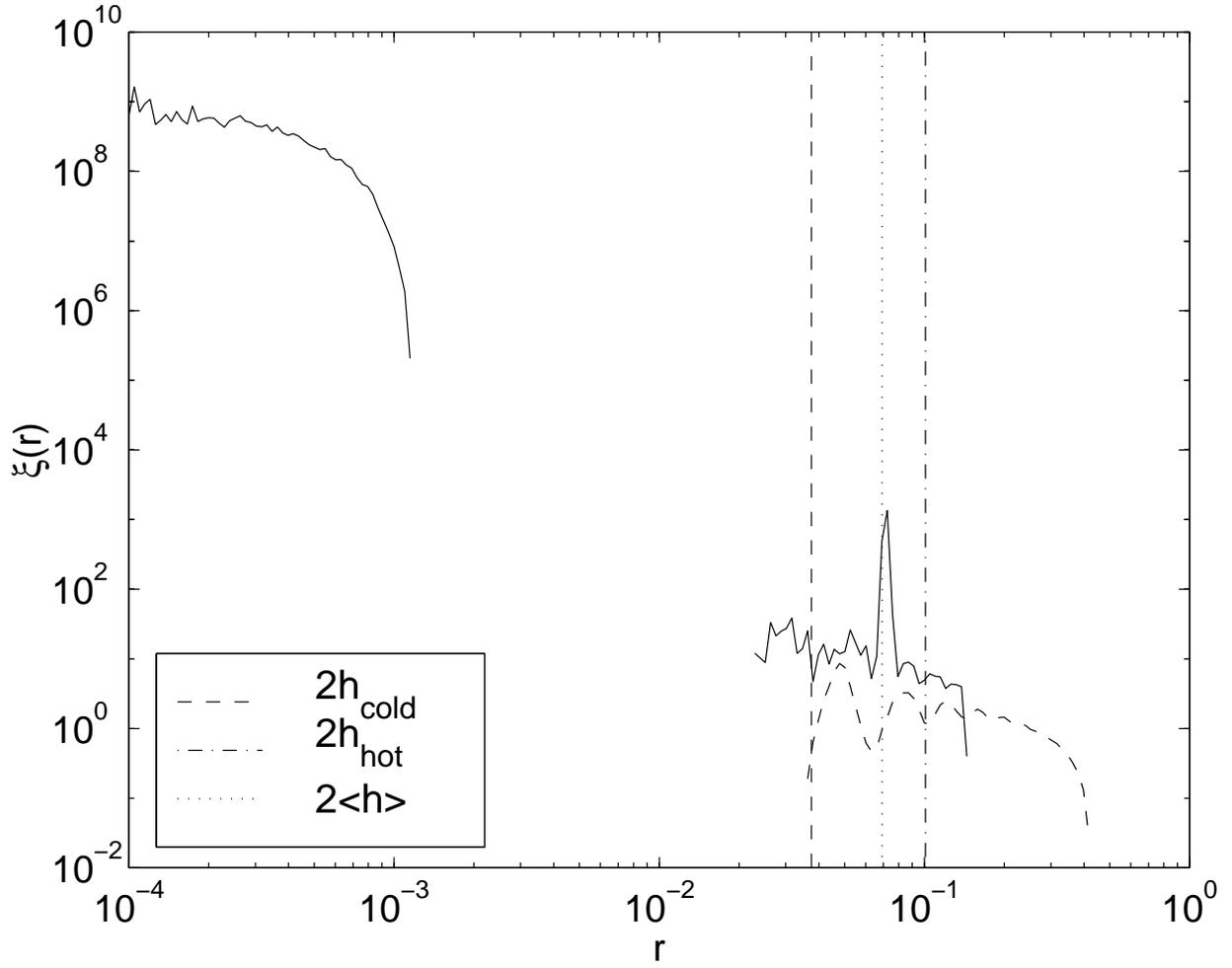}
\caption{The 2-point correlation function about the cold clump.
Marked are the smoothing limits for the particles of the cold clump (dashed), the hot gas (dot-dashed), and the average (dotted) which is the value used in these tests for a hot particle interacting with the cold clump.  For comparison, the 2-point correlation function for the hot gas away from the cold clump is given by the dashed curve.}
\label{fig.Results.CorrFunction_2001}
\end{figure}

\begin{figure}
\epsscale{\figscale}
\plotone{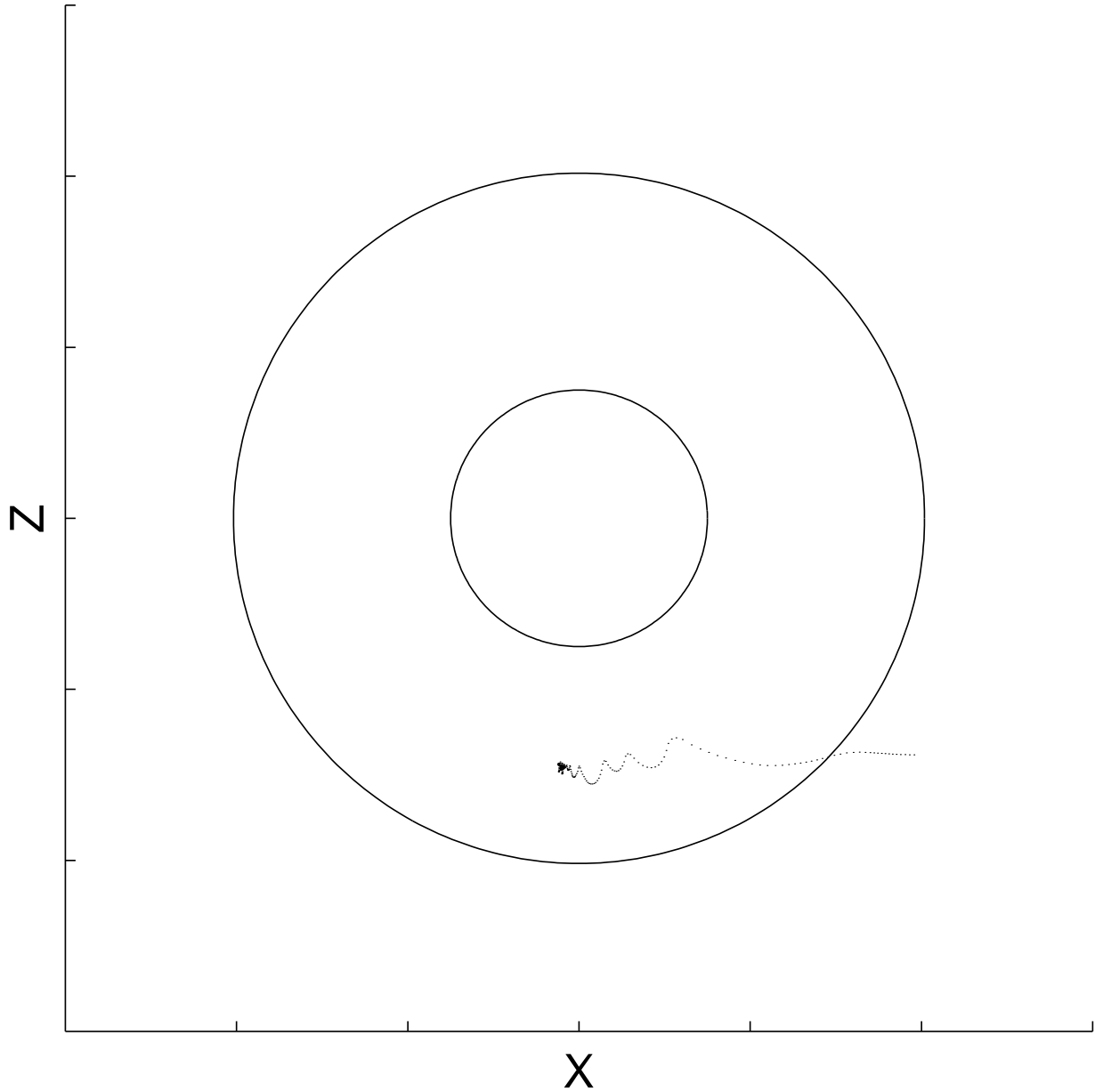}
\caption{Path of a hot particle onto a cold, Mach \onethird\ clump.
The position of the particle at successive iterations is plotted relative to the cold clump.  Flow is from the right (the clump is moving left).  Again, the outer circle has a radius of $2h_{hot}$ while the inner radius is $2h_{clump}$.}
\label{fig.Results.Flow_1439}
\end{figure}

\begin{figure}
\epsscale{\figscale}
\plotone{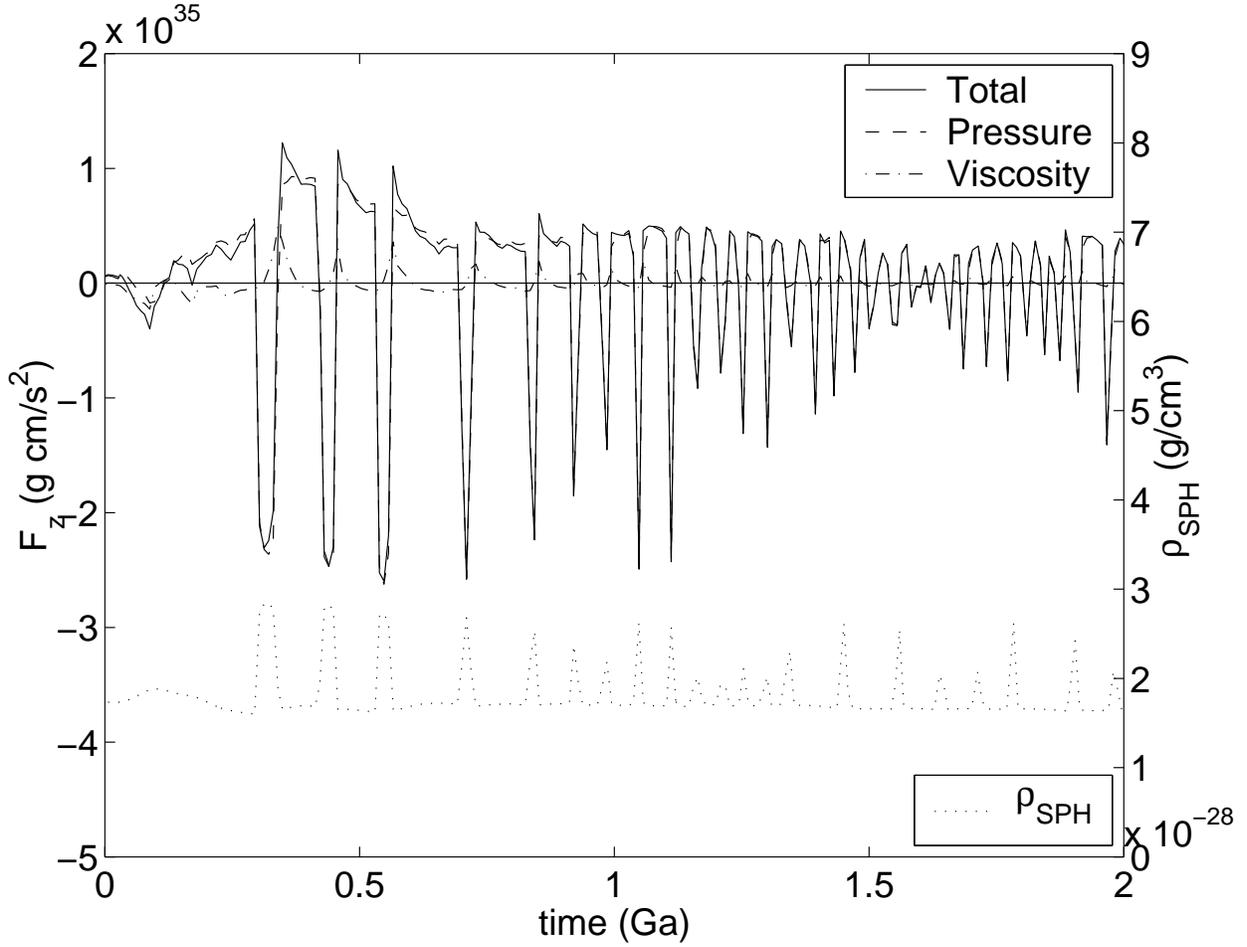}
\caption{The forces on a remora particle that propel it in a direction perpendicular to the direction of travel.
Plotted is the total force (solid line) as well as the contributions from the gas pressure (dashed line) and the artificial viscosity (dash-dotted line).  The gas pressure provides the bulk of the instantaneous forcing and hence frequently overlaps the total force.  Positive forces are directed toward the clump centre. At the bottom is plotted the gas density calculated by the SPH routine (dotted line).  The axis to the right relates to this plot.}
\label{fig.Results.ZForces_1439}
\end{figure}

\begin{figure}
\epsscale{\figscale}
\plotone{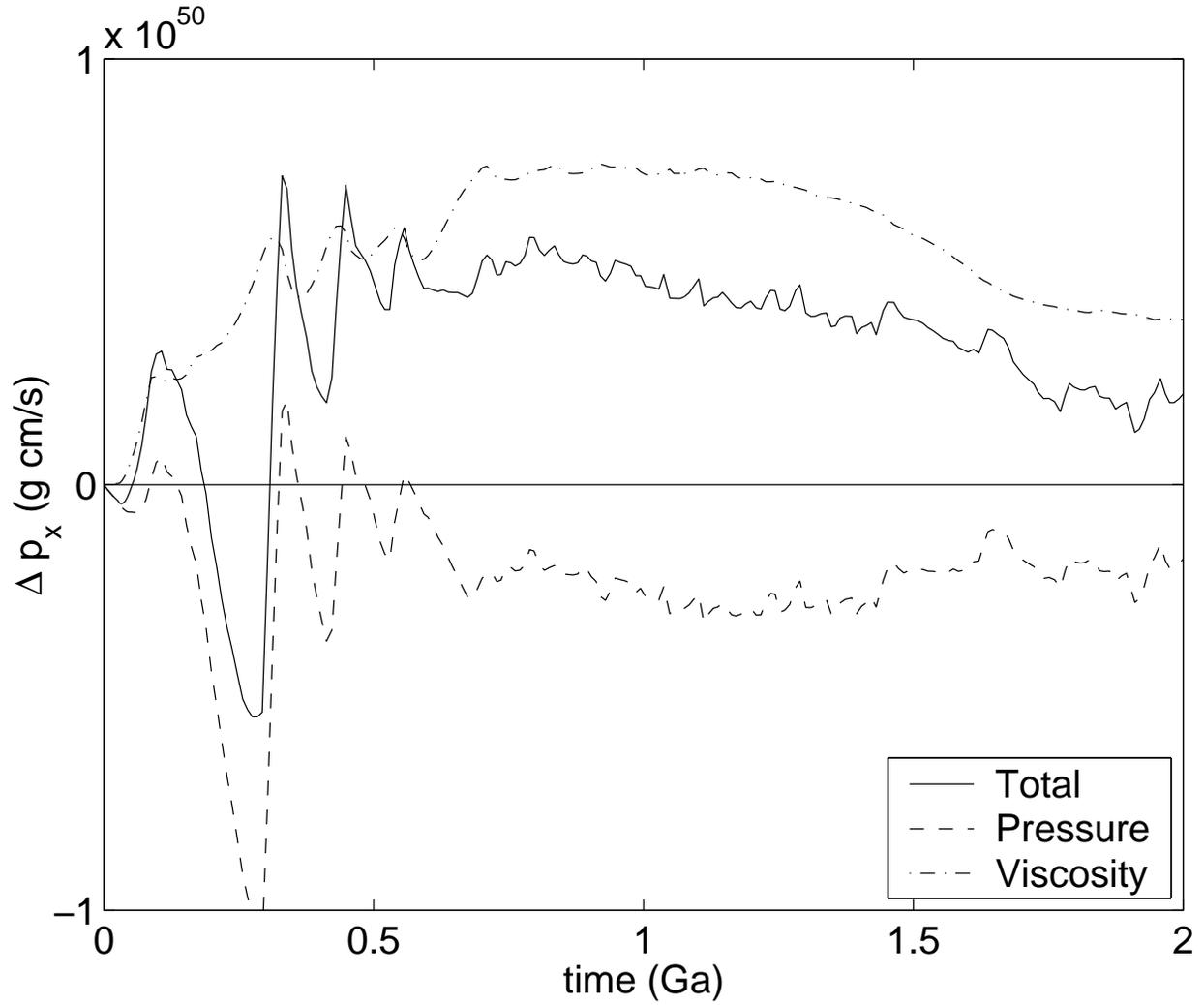}
\caption{The cumulative momentum gained by a remora particle due to SPH forces.
The panel plots the forces gained in the direction of travel of the cold clump as it accumulated with time. Plotted is the total momentum boost (solid line) as well as the contributions of the SPH pressure term (dashed line) and the artificial viscosity (dash-dotted line).}
\label{fig.Results.CumulativeMom_1439}
\end{figure}

\end{document}